\documentclass[preprint,12pt]{elsarticle}

\usepackage{amssymb}

\newcounter{bla}

\usepackage[colorlinks=true,linkcolor=blue,citecolor=blue]{hyperref} 
\usepackage{amsmath}
\usepackage{lipsum}
\usepackage{amsfonts}
\usepackage{graphicx}
\usepackage{epstopdf}
\usepackage{caption}
\usepackage{subcaption}

\usepackage{floatrow}
\usepackage{algorithm}
\usepackage[noend]{algpseudocode}
\usepackage{orcidlink}

\usepackage{wrapfig}
\ifpdf
  \DeclareGraphicsExtensions{.eps,.pdf,.png,.jpg}
\else
  \DeclareGraphicsExtensions{.eps}
\fi

\def\Bb {{\bf B}}
\def\Db {{\bf D}}
\def\Eb {{\bf E}}

\def\Hb {{\bf H}}
\def\Jb {{\bf J}}

\journal{Computer Physics Communications}

\begin{document}

\begin{frontmatter}

\title{Two-Fluid Physical Modeling of Superconducting Resonators in the ARTEMIS Framework}

\author[a]{Revathi Jambunathan\corref{author}\orcidlink{0000-0001-9432-2091}}
\author[a]{Zhi Yao \orcidlink{0000-0001-5863-8275} }
\author[b]{Richard Lombardini \orcidlink{0000-0002-0621-6131}}
\author[b]{Aaron Rodriguez}
\author[a]{Andrew Nonaka \orcidlink{0000-0003-1791-0265}}

\cortext[author] {Corresponding author.\\\textit{E-mail address:} rjambunathan@lbl.gov}
\address[a]{Center for Computational Sciences and Engineering, Lawrence Berkeley National Laboratory}
\address[b]{Department of Physics and Environmental Science, St. Mary's University (San Antonio, TX)}

\begin{abstract}
In this work, we implement a new London equation module for superconductivity in the GPU-enabled ARTEMIS framework, and couple it to a finite-difference time-domain solver for Maxwell's equations.
We apply this two-fluid approach to model a superconducting coplanar waveguide (CPW) resonator. 
We validate our implementation by verifying that the theoretical skin depth and reflection coefficients can be obtained for several superconductive materials, with different London penetration depths, over a range of frequencies.
Our convergence studies show that the algorithm is second-order accurate in both space and time, except at superconducting interfaces where the approach is spatially first-order.
In our CPW simulations, we leverage the GPU scalability of our code to compare the two-fluid model to more traditional approaches that approximate superconducting behavior and demonstrate that superconducting physics can show comparable performance to the assumption of quasi-infinite conductivity as measured by the Q-factor.
\end{abstract}

\begin{keyword}
Superconducting Materials \sep Maxwell's Equations; London Equations; Finite-Difference Time-Domain; Two-Fluid Model; Resonators; Microelectronics

\MSC 35-04 \sep 35Q60 \sep 78M20 \sep 82D55
\end{keyword}

\end{frontmatter}

\section{Introduction}
\label{sec:Introduction}
 
The tremendous growth in materials research as well as the race to miniaturize microelectronic devices has accelerated the adoption and integration of novel materials in traditional CMOS devices.
Superconducting materials exhibit unique characteristics, such as, greatly reduced loss compared to metals, the expulsion of magnetic fields via the Meissner effect \cite{meissner1933}, quantum tunneling effects, and flux quantization \cite{tinkham1974electromagnetic}, making them a promising candidate to produce high-fidelity and high-coherence circuits. While our focus is on microelectronics, applications of these materials also extend to imaging \cite{kirtley2010fundamental}, tokamaks \cite{gryaznevich2013progress}, accelerators \cite{rossi2012superconducting,padamsee2001science}, and magnetic levitation \cite{moon2008superconducting,ma2003superconductor}. 
In microelectronics, these materials are used in resonators, qubits with Josephson junctions to build quantum devices, circuit quantum electrodynamics devices (cQED) \cite{burkard2020superconductor,wallraff2004strong}, and in superconducting quantum interference devices (SQUID). Resonators are devices where the measured field attains maximum amplitude at a designed resonant frequency. Superconducting coplanar waveguide (CPW) resonators are used in quantum computing applications as they are ideal for control and readout \cite{caputo2021abelian,sage2011study} and as an interface between resonators and qubits.

In order to design and optimize such devices without resorting to expensive trial-and-error fabrication and measurement cycles, we require an accurate simulation tool that can model the superconducting behavior over a wide range of frequencies for a given configuration of material properties.
In this work, we are interested in a classical description of the superconducting materials to investigate the interaction of the electromagnetic signals with the superconducting sub-components of the CPW.
Traditionally, these sub-components have been approximated as a perfect conductor or as a highly conductive material or the interaction is reduced to an empirical model with an resistance-inductance-capacitance (RLC) response to the incoming signal \cite{goppl2008coplanar}.
However, such methods do not capture the non-linear coupling between electromagnetic and superconducting physics and more accurate numerical descriptions are required.

For accurate classical descriptions of superconducting materials, the London equations \cite{london1935electromagnetic} provide foundational constitutive relationships. These equations are coupled with Maxwell's equations to fully describe the 
interaction between electromagnetic fields and currents in superconducting materials.
In the past decades, there has been interest in incorporating the London equations into the widely-used finite-difference time-domain (FDTD) approach \cite{kunz1993finite} for solving Maxwell's equations.
This two-fluid approach has been derived independently from two different mathematical formulations, however, they both lead to functionally identical numerical implementation.
The first mathematical formulation involves the use of a complex conductivity to describe the contribution from superconductivity. 
In Rittweger et al.~\cite{rittweger1992finite} and similar work by others \cite{xiao1994extended,xiao19953d,hofschen1996improvements}, the complex conductivity was incorporated into a frequency-domain representation of Maxwell's equations and converted to the equivalent time-domain representation.  This resulted in the inclusion of an additional source term, equal to the time-integral of the electric field, in Ampere's law. 
The second mathematical formulation involves the use of a two-fluid model \cite{Gorter34,van1981principles}, where, the total current density is the sum of the standard conductive current plus a superconducting current whose evolution is governed by the first London equation.
This approach, used by a number of works \cite{megahed1995nonlinear,okazaki1999superconducting,mollai2013analysis,li2021unconditionally,li2022extended}, is analytically equivalent to that of the first, since the superconducting current is in fact, a scaled time-integral of the electric field.
Alternative approaches have also been considered, such as by Yun et al.~\cite{Shen-Yun10}, where a shift-operator technique is incorporated into the FDTD framework to directly account for a complex conductivity.

The main challenge in using a coupled explicit Maxwell-London solver for CPW resonators is the large disparity in length-scale of the London penetration depth, (typically 10-400~nm), and the size of the CPW resonator ($\sim$ 1000)~$\mu$m. As a result of the explicit time integration, the disparity in the temporal scales required to resolve the speed of light and that required for the low frequency signal to achieve resonance results in simulations that can require $10^6$ time steps or more. Using traditional CPU-based solvers render such simulations impossible to perform and a scalable GPU-enabled code is required. Therefore, we use our GPU-enabled open source framework, called ARTEMIS \cite{yao2022massively} developed to model electromagnetic signals in microelectronics devices. 

In this paper, we describe our implementation of the two-fluid approach in ARTEMIS for modeling interactions between electromagnetic signals and superconducting components and apply it to the study of CPW resonators. The GPU speedup and scalability of our code allows for rigorous validation and case studies with frequencies comparable to operating conditions of devices, that would not be possible otherwise.
The rest of this paper is organized as follows.
In Section \ref{sec:model}, we describe our two-fluid model to couple Maxwell and London equations.
In Section \ref{sec:implementation}, we describe our numerical method and implementation in the ARTEMIS framework.
In Section \ref{sec:validation}, we present the skin depth and reflection coefficient analysis, and validate that we are able to reproduce theoretically-predicted behavior.  We also present spatial and temporal convergence tests for a number of material configurations.
Finally, in Section \ref{sec:coplanar}, we perform simulations of a CPW resonator and compare the results from our two-fluid model to those obtained from a simpler, purely-conductive approximation for superconductivity that can be accomplished with a standard FDTD Maxwell solver.

\section{Two-Fluid Model for Superconductivity}
\label{sec:model}
The thermodynamic model proposed by Gorter and Casimir \cite{Gorter34} states that superconducting materials at temperatures between absolute zero and the critical temperature will contain both conductive and superconductive currents, hence the term ``two-fluid model'', where one includes normal electrons (with finite conductivity, $\sigma > 0$) and the other superconducting electrons, i.e., Cooper pairs \cite{Bardeen57}.  According to this model, at $T=0$~K, all electrons condense to Cooper pairs and $\sigma = 0$, leading to pure superconducting behavior. To model such behavior, we first begin with the full-form of the dynamic Maxwell's equations, i.e., Ampere and Faraday's laws,
\begin{equation}
\nabla\times\Hb = \Jb + \frac{\partial\Db}{\partial t},\label{eq:Ampere}
\end{equation}
\begin{equation}
\nabla\times\Eb = -\frac{\partial\Bb}{\partial t}\label{eq:Faraday}
\end{equation}
where, $\Db = \epsilon\Eb$ is the electric displacement, $\Eb$ is the electric field, $\Bb = \mu\Hb$ is the magnetic flux density, and $\Hb$ is the magnetic field.
The permittivity of the medium, $\epsilon$, is the product of vacuum permittivity, $\epsilon_0$, and the dimensionless relative permittivity, $\epsilon = \epsilon_0\epsilon_r$.
Similarly, the permeability of the medium, $\mu$, is the product of vacuum permeability, $\mu_0$ and the unit-less relative permeability, $\mu = \mu_0\mu_r$.
Consistent with the model proposed by Gorter and Casimir\cite{Gorter34}, in the two-fluid model, the total electric current density, $\Jb$, in equation (\ref{eq:Ampere}), is given by the sum of the conductive current, $\sigma\Eb$, and the superconducting current, $\Jb_{s}$, such that, $\Jb=\sigma\Eb + \Jb_{s}$, with conductivity $\sigma$.

In order to obtain the superconducting current, we invoke the classical model for superconductivity given by the London equations.
The first London equation, given below, can be derived by combining the Lorentz force with the unbounded acceleration of electrons in the presence of an electric field.
\begin{equation}
    \frac{\partial\Jb_s}{\partial t} = \frac{n_s e^2}{m}\Eb,\label{eq:London1}
\end{equation}
where, $n_s$ is the number density of superconducting electrons, $m$ is the electron mass, and $e$ is the elementary charge.
If we define  $\lambda = \sqrt{m/(n_s e^2\mu)}$ as the London penetration depth, Equation (\ref{eq:London1}) can be written as
\begin{equation}
    \frac{\partial\Jb_s}{\partial t} = \frac{1}{\lambda^2\mu}\Eb.\label{eq:London1a}
\end{equation}
The typical range for London penetration depth of superconducting materials is $\mathcal{O}(10-100)$~nm.
As previously stated, a superconducting material may still exhibit finite conductivity that reduces to zero as the temperature approaches absolute zero.
Thus, to model such systems, we require the two-fluid approach, where Maxwell's equations given in equations (\ref{eq:Ampere}) and (\ref{eq:Faraday}) are coupled with the first London equation given in equation (\ref{eq:London1}) to provide a fully classical description of the superconducting physics.

\section{Numerical Method and Implementation}\label{sec:implementation}
We employ the standard Yee grid configuration for electrodynamics, where, the normal components of $\Bb$ fields are defined on cell-faces, and the tangential components of $\Eb$ are defined on the cell-edges.
The standard explicit FDTD scheme for Maxwell's equations on a Yee grid uses a leap-frog discretization in time, where $\Eb$ is updated at integer time levels and $\Bb$ is updated at half-integer time levels.
In the superconducting regions within the domain, the superconducting current, $\Jb_s$ are defined using the same spatial discretization as the electric field, i.e., tangential currents on the cell-edges, and same temporal discretization as the magnetic field.
\textcolor{black}{We note that this algorithm is constrained by the fact that the interface between a non-superconducting and superconducting material is always grid aligned.  Thus, all cell-edges that lie on such interfaces are considered superconducting and contain tangential components of $\Jb_s$.
Also, due to the spatial Yee grid and leap-frog temporal discretization, the numerical scheme is second-order in space and time.  If there is a sharp discontinuity in either the conductivity or the inverse of the penetration depth, the algorithm is first-order in space; for cases where the conductivity is smoothly varying and at superconducting interfaces the inverse penetration depth smoothly varies, second-order spatial convergence is retained.  It should also be noted that the case of non-superconducting material implies that $1/\lambda\rightarrow 0$.}
The integration scheme that we implemented in ARTEMIS is described below in Algorithm~\ref{alg:TwoFluid} and it is analytically equivalent to a standard leap-frog approach. For diagnostic purposes, we split the time-update of the magnetic field, $\Bb$, and current density, $\Jb_s$, into two half timestep updates.
In Algorithm~\ref{alg:TwoFluid}, we describe the steps to advance the solution from time level, $t^n$ to the next time level at $t^n + \Delta t$, where the superscript on a variable indicates the time step index.

\begin{algorithm}[!htb]
\caption{Two-fluid Maxwell London Algorithm}\label{alg:TwoFluid}
\begin{algorithmic}[1]
\State Integrate $\Bb^{n} \rightarrow \Bb^{n+1/2}$ using $\Eb^{n}$ and Faraday's law (\ref{eq:Faraday}) (first half of B-update)
\begin{equation*}
\Bb^{n+1/2} = \Bb^n - \frac{\Delta t}{2}(\nabla\times\Eb^n).
\end{equation*}

\State In superconducting regions, integrate $\Jb_s^n \rightarrow \Jb_s^{n+1/2}$, (first half of J-update)
\begin{equation*}
    \frac{\Jb_s^{n+1/2} - \Jb_s^n}{\Delta t/2} = \frac{1}{\lambda^2\mu}\Eb^n
\end{equation*}
\State Integrate $\Eb^n \rightarrow \Eb^{n+1}$ using $\Bb^{n+1/2}$, and $\Jb^{n+1/2} \equiv \sigma[(1-\theta)\Eb^n + \theta\Eb^{n+1}] + \Jb_s^{n+1/2}$ in Ampere's law (\ref{eq:Ampere}).
Here, $\theta$ is a parameter that controls the temporal discretization of the conductive current, where $\theta=1$ is a first-order backward-Euler discretization and $\theta=0.5$ is a second-order time-centered discretization. The electric field update is given by
\begin{equation*}
\begin{aligned}
    \Eb^{n+1} = \left(1 + \theta\frac{\sigma\Delta t}{\epsilon}\right)^{-1} \bigg( & \Eb^n + \frac{\Delta t}{\epsilon}\nabla\times \Hb^{n+1/2} \\
   &  -(1-\theta)\frac{\sigma\Delta t}{\epsilon}\Eb^n - \frac{\Delta t}{\epsilon}\Jb_s^{n+1/2} \bigg)
\end{aligned}
\end{equation*}
For the convergence testing simulations, we use $\theta=0.5$; the remainder of the simulations use $\theta=1$.
\State In superconducting regions, integrate $\Jb_s^{n+1/2} \rightarrow \Jb_s^{n+1}$ (second half of J update)
\begin{equation*}
    \frac{\Jb_s^{n+1} - \Jb_s^{n+1/2}}{\Delta t/2} = \frac{1}{\lambda^2\mu}\Eb^{n+1}
\end{equation*}
\State Integrate $\Bb^{n+1/2} \rightarrow \Bb^{n+1}$ using $\Eb^{n+1}$ in Faraday's law (\ref{eq:Faraday}) (second half of B-update)
\begin{equation*}
\Bb^{n+1} = \Bb^{n+1/2} - \frac{\Delta t}{2}(\nabla\times\Eb^{n+1}).
\end{equation*}
\end{algorithmic}
\end{algorithm}

For domain boundary conditions, the code includes the standard options for periodic, perfect electric conductor (PEC), and perfectly matched layer (PML) \cite{berenger1996three,shapoval2019two}.
We note that the interaction between a London region and PML is not well-understood.
Thus, in our simulations, we use a domain large enough such that the signal does not interact with London regions close to the PML boundaries, or include an air gap in-between the London region and the domain boundary making sure the overall characterization of the device is not significantly affected.

ARTEMIS is built on the AMReX framework for block-structured mesh calculations \cite{zhang2021amrex} and leverages many of the computational kernels from the ECP-funded electromagnetic Particle-In-Cell WarpX \cite{vay2018warp} code.
Thus, the ARTEMIS code is portable and scalable to the largest multicore and GPU-based supercomputers.
\textcolor{black}{We note that all of the simulations in this paper are performed using uniformly-sized cuboid cells and leverage the efficient and scalable MPI+CUDA implementation provided by AMReX and WarpX.  More specifically, we use a hierarchical parallelization model where the domain is divided into boxes that are distributed to MPI ranks, and computational work is performed by distributing individual grid cells to GPU threads.}
Using three NERSC HPC systems (Perlmutter GPU partition, Perlmutter CPU partition, haswell CPU partition),
we find that on a node-by-node basis, the perlmutter GPU partition offers a 10.5x speedup compared to Perlmutter CPU partition, and a 56x speedup over the haswell CPU partition.
We have recently demonstrated the near-perfect weak scaling performance of the code on up to 2,000 GPUs \cite{yao2022massively,Sawant2022} and due to the explicit nature of the algorithm, the scaling properties of the new London module will behave the same.

\section{Physical and Numerical Validation}\label{sec:validation}

In this section, we first validate our implementation by examining the skin depth within superconducting material and measuring the reflection properties at superconducting interfaces and comparing the results with theoretical predictions.
We then demonstrate the numerical convergence properties of our implementation. 

\subsection{Skin Depth}
\label{sec:skindepth}

The general formula for the skin depth in a normal conductor is well-known \cite{griffiths2005introduction}. 
Here we derive the analogous expression that includes the superconducting current.
Consider the Maxwell London model in a homogeneous medium with uniform $\epsilon$, $\mu$, $\sigma$, $\lambda$, and no free charges.
Applying the curl to Equations  (\ref{eq:Ampere}) and (\ref{eq:Faraday}) using $\nabla\cdot\Bb = \nabla\cdot\Eb = 0$, we get,
\begin{eqnarray}
\nabla^2 \bf{E} & = & \mu \epsilon \frac{\partial^2 \bf{E}}{\partial t^2} + \mu \sigma \frac{\partial \bf{E}}{\partial \it{t}} + \frac{\bf{E}}{\lambda^2}\\
\nabla^2 \bf{B} & = & \mu \epsilon \frac{\partial^2 \bf{B}}{\partial t^2} + \mu \sigma \frac{\partial \bf{B}}{\partial \it{t}} + \frac{\bf{B}}{\lambda^2}.
\end{eqnarray}
These equations admit plane wave solutions. Let us consider a plane wave traveling along the $z$-direction given by
\begin{eqnarray}
\label{Esol}
{\bf E}(z,t) &=& {\bf E}_0 \it{e}^{\it{i(kz-\omega t)}} \\
\label{Bsol}
{\bf B}(z,t) &=& {\bf B}_0 \it{e}^{\it{i(kz-\omega t)}} 
\end{eqnarray}
where, \textcolor{black}{$\bf E_0$} and \textcolor{black}{$\bf B_0$} are the magnitude of the electric and magnetic fields, $\omega$ is the frequency, and the wavenumber, $k$, is complex and equal to
\begin{equation}
\label{eq:nk}
k = \sqrt{\bigg(\mu \epsilon \omega^2 - \frac{1}{\lambda^2} \bigg) + i \mu \sigma \omega}.
\end{equation}
Re-writing the complex wavenumber as $k=\gamma + i\kappa$ and taking the square root of the complex term in Eq.~\ref{eq:nk}, we can write the real and imaginary part of the wavenumber as,
\begin{equation}
\label{ngamma}
\gamma = \sqrt{\frac{\sqrt{\bigg(\mu \epsilon \omega^2 - \frac{1}{\lambda^2}\bigg)^2 + (\mu \sigma \omega)^2} + \bigg(\mu \epsilon \omega^2 - \frac{1}{\lambda^2}\bigg)}{2}}
\end{equation}
\begin{equation}
\label{nkappa}
\kappa = \sqrt{\frac{\sqrt{\bigg(\mu \epsilon \omega^2 - \frac{1}{\lambda^2}\bigg)^2 + (\mu \sigma \omega)^2} - \bigg(\mu \epsilon \omega^2 - \frac{1}{\lambda^2}\bigg)}{2}}.
\end{equation}
The skin depth, $\delta$, for the superconductor is simply given by
\begin{equation}
\delta = \frac{1}{\kappa}.\label{eq:skin}
\end{equation}
Note that when the superconducting current is not included (i.e., $1/\lambda^2=0$), the skin depth reduces to the well-known expression for a conductor.
Also, in the limit as $\omega\rightarrow 0$, the skin depth reduces to the London penetration depth, $\lambda$.

\subsubsection*{Comparison of skin depth with theory}
We now measure the skin depth using a series of tests and compare against theoretical predictions.
The computational domain in these tests are homogeneous consisting of superconducting metal with a London penetration depth, $\lambda=400$~nm, vacuum permittivity and permeability, and uniform conductivity. 
To study the effect of conductivity on skin depth and compare with theory, we perform tests with three different values, i.e., $\sigma=0, 10^4$, and $10^7$ S/m.
For each value of conductivity, we perform tests with four different frequencies, $f=25$~GHz, 100~GHz, 1~THz, and 100~THz,
In each case, we perform one-dimensional simulations with $\Delta z=10$~nm and a domain extending from $-L_z$ to $L_z$. The value of $L_z$ is dependent on the frequency and is chosen to be long enough such that the signal does not interact with the domain boundaries. 
Thus, we use  $L_z = 128$~mm, 32~mm, 3.2~mm, and 32~$\mu$m, respectively, for the four frequency values given above.
We excite the system with an electric field given by $E_y = \sin{(2\pi f)}$ at the center of the domain, $z=0$, and measure the peak amplitude of the signal at the source, i.e., at $z=0$ and at an observation point along the propagation direction at $z_0>0$.
These measurements will be compared against the theoretical skin-depth.
In each simulation, we choose the observation point, $z_0$, such that it lies on a grid-point closest to theoretical skin depth for that particular configuration of conductivity and frequency used in the simulation.
We use a CFL of 0.9, with a corresponding time step of $\Delta t = 0.03$~fs and run each simulation to $t = 5T$~s, where $T=1/f$ is the period of the excitation.

To measure the skin depth from our simulations, we use
\begin{equation}
\delta_{\rm computed}^{\rm supercond} = \frac{z_0}{\ln({E^{\rm peak}_{z=0}/E^{\rm peak}_{z=z_0}})},
\end{equation}
where, the $E^{\rm peak}$ values correspond to the final peak amplitude of the measured signal during the simulation.
In Table~\ref{tab:skin1}, we compare the theoretical values of the skin depth ($\delta_{\rm theory}^{\rm supercond}$) and the computed skin depth ($\delta_{\rm computed}^{\rm supercond}$) in columns 4 and 5, respectively.
We obtain excellent agreement between the theoretical and computed skin depths using the two-fluid approach implemented in ARTEMIS.
We note that, as predicted by the theory, for all values of conductivity, the skin depth approaches the London penetration depth as we decrease the frequency. 
On the other hand, as the frequency increases, the skin depth either increases or decreases depending on the conductivity.
For reference, we also include in column~3, the theoretical skin depth assuming the metal is conductive with no superconducting behavior ($\delta_{\rm theory}^{\rm cond}$) to highlight the difference in the physical behavior of the two models indicating the effect on the simulation if superconductivity is not accounted for.

\begin{table}[tb!]
\centering
\begin{tabular}{p{1.8cm} p{2.1cm} p{2.3cm} p{2.5cm} p{2.5cm} }
\hline
$\sigma$ [S/m] & $f$ & $\delta_{\rm theory}^{\rm cond}$ [nm] & $\delta_{\rm theory}^{\rm supercond}$ [nm] & $\delta_{\rm computed}^{\rm supercond}$ [nm]\\
\hline
0 & 25 GHz & $\infty$ & 400 & 400 \\
0 & 100 GHz & $\infty$ & 400 & 400 \\
0 & 1 THz & $\infty$ & 400 & 400 \\
0 & 100 THz & $\infty$ & 734 & 739 \\
\hline
$10^4$ & 25 GHz & 50331 & 400 & 400 \\
$10^4$ & 100 GHz & 15920 & 400 & 400 \\
$10^4$ & 1 THz & 5047 & 400 & 400 \\
$10^4$ & 100 THz & 656 & 448 & 450 \\
\hline
$10^7$ & 25 GHz & 1592 & 399 & 395 \\
$10^7$ & 100 GHz & 503 & 350 & 350 \\
$10^7$ & 1 THz & 159 & 153 & 153 \\
$10^7$ & 100 THz & 16 & 16 & 13 \\
\hline
\end{tabular}
\caption{Comparison of theoretical and computed skin depths as a function of $\sigma$ and $f$ for a superconducting material with London penetration depth, $\lambda=400$~nm.}
\label{tab:skin1}
\end{table}

\subsection{Reflection Coefficient}
Next, we examine the reflection coefficient, $R$, as a function of frequency, $\omega$, for a signal propagating from vacuum medium at normal incidence to a conductor or superconductor. 
With the same plane waves, previously discussed in Section \ref{sec:skindepth}, the signals travel in the positive $z$ direction resulting in transmission and reflection.
The incident $I$, reflected $R$, and transmitted $T$ waves are given as
\begin{eqnarray}
\bf{E}_{\it{I}}(\it{z},\it{t}) & = & E_{0,I} e^{i(k_1 z - \omega t)} \bf{\hat{y}} \\
\bf{B}_{\it{I}}(\it{z},\it{t}) & = & -\frac{E_{0,I}}{v_1} e^{i(k_1 z - \omega t)} \bf{\hat{x}} \label{eq:Bi}\\
\bf{E}_{\it{R}}(\it{z},\it{t}) & = & E_{0,R} e^{i(-k_1 z - \omega t)} \bf{\hat{y}} \\
\bf{B}_{\it{R}}(\it{z},\it{t}) & = & \frac{E_{0,R}}{v_1} e^{i(-k_1 z - \omega t)} \bf{\hat{x}} \label{eq:Br}
\end{eqnarray}
\begin{eqnarray}
\bf{E}_{\it{T}}(\it{z},\it{t}) & = & E_{0,T} e^{i(k_2 z - \omega t)} \bf{\hat{y}} \\
\bf{B}_{\it{T}}(\it{z},\it{t}) & = & -\frac{E_{0,T}}{v_2} e^{i(k_2 z - \omega t)} \bf{\hat{x}} . \label{eq:Bt}
\end{eqnarray}
Let us label the vacuum medium as medium 1 with wave speed $v_1=\omega/k_1$, where the wavenumber, $k_1=\omega \sqrt{\epsilon_1 \mu_1}$, and medium 2 is the conductor or superconductor with wave speed $v_2 = \omega/k_2$, where $k_2$ is given by Equation (\ref{eq:nk}).  

The parallel $\bf{E}$ and normal $\bf{H}$ fields are continuous at an interface, i.e.,
\begin{eqnarray}
\label{Ebound}
\bf{E}_{\text{1}}^{||} & = & \bf{E}_{\text{2}}^{||} \\
\label{Bbound}
\frac{\bf{B}_{\text{1}}^{\perp}}{\mu_1} & = & \frac{\bf{B}_{\text{2}}^{\perp}}{\mu_2}
\end{eqnarray}
If the interface is at $z=0$, then Equation (\ref{Ebound}) implies
\begin{equation}
E_{0,I}+E_{0,R} = E_{0,T}, \label{eq:Etan}
\end{equation}
and Equation (\ref{Bbound}) combined with Equations~\ref{eq:Bi}, \ref{eq:Br}, and \ref{eq:Bt} implies
\begin{equation}
\frac{1}{\mu_1 v_1}(E_{0,I}-E_{0,R}) = \frac{E_{0,T}}{\mu_2 v_2},
\end{equation}
or equivalently,
\begin{equation}
E_{0,I}-E_{0,R} = \sqrt{\frac{\mu_1}{\epsilon_1}} \frac{k_2}{\mu_2 \omega} E_{0,T}. \label{eq:Btan}
\end{equation}
Combining Equations (\ref{eq:Etan}), (\ref{eq:Btan}), and (\ref{eq:beta}), we obtain the following relationships between the incident amplitude and both the reflected and transmitted
\begin{eqnarray}
E_{0,R}=\frac{1-\beta}{1+\beta}E_{0,I} \\
E_{0,T}=\frac{2}{1+\beta}E_{0,I}.
\end{eqnarray}
where, $\beta$ is a complex term given by,
\begin{equation}
\beta = \sqrt{\frac{\mu_1}{\epsilon_1}} \frac{k_2}{\mu_2 \omega}. \label{eq:beta}
\end{equation}
Thus, the theoretical reflection coefficient is
\begin{equation}
R = \left | \frac{E_{0,R}}{E_{0,I}} \right |^2 
= \left | \frac{1-\beta}{1+\beta} \right|^2. 
\label{eq:Ref}
\end{equation}
An interesting result of this analysis is that for the cases where $\sigma=0$, there is a cut-off frequency below which the reflection coefficient $R=1$, with a dropoff in $R$ greater than this cutoff.
Specifically, if $\sigma=0$, then Eq.~\ref{eq:nk} reduces to
\begin{equation}
    k_2 = \sqrt{\mu_2 \epsilon_2 \omega^2 - \frac{1}{\lambda^2}},
\end{equation}
and therefore,
\begin{equation}
    \beta = \sqrt{\frac{\mu_1}{\epsilon_1}} \frac{1}{\mu_2 \omega}\begin{cases}
    i \sqrt{\frac{1}{\lambda^2} - \mu_2 \epsilon_2 \omega^2} &\omega \leq \frac{1}{\lambda \sqrt{\mu_2 \epsilon_2}} \\
    \sqrt{\mu_2 \epsilon_2 \omega^2 - \frac{1}{\lambda^2}} &\omega > \frac{1}{\lambda \sqrt{\mu_2 \epsilon_2}} 
    \end{cases} .\label{eq:beta_cases}
\end{equation}
Substituting Equation (\ref{eq:beta_cases}) in Equation (\ref{eq:Ref}), we get complete reflection, i.e., $R=1$, for $\omega \leq 1/(\lambda \sqrt{\mu_2 \epsilon_2})$.
For finite conductivity ($\sigma>0$), $R$ smoothly decreases from 1 to 0 with increasing $\omega$.
Next, we will perform demonstration tests to validate our implementation and highlight these key features of the reflection coefficient.

\subsubsection*{Comparison of Reflection Coefficient with Theory}
To compare the reflection coefficient with theoretical predictions, we perform one-dimensional simulations from $z=0$ to $z=89.6~\mu$m.
For $z < 7~\mu$m we define a vacuum region (medium 1) and for $z >= 7~\mu$m we define a superconducting region with $\lambda=100$~nm (medium 2).
We perform two tests with different values for conductivity, $\sigma=$0 and $10^4$~S/m, and use vacuum permittivity and permeability everywhere.
We use periodic boundary conditions in $x$ and $y$, and PML at the low and high $z$ boundaries. Note that, we terminate the simulation before the signal interacts with the PML boundary.
We discretize the domain with a uniform mesh with 4,480 grid cells such that $\Delta z = 20$~nm; thus we resolve the London penetration depth sufficiently in our validation tests.
We then excite the system at $z=100$~nm with a frequency-modulated Gaussian pulse given by,
\begin{equation}
    E_y = \exp{  \frac{-(t-t_o)^2}{2t_w^2}}
\end{equation}
where, $t_w=1/(2f)$ is the Gaussian width of the pulsar, $t_o=4t_w$ is the initial pulse duration, and the frequency, $f$, is 100~THz. 
For the cases we present here, we use CFL=0.95 resulting in a timestep of 0.0632 fs, and the simulations were run for 15,800 timesteps, such that we resolve one-hundredth of the frequency of the input signal. 
To compute the reflection, we first measure the total signal just upstream of the interface, at $z=6.8~\mu$m, from the simulation. 
We then measure the input signal at the same observation point from a separate simulation, performed using the same simulation parameters, but, with pure vacuum conditions throughout the simulation domain without any interface. 
To obtain the reflected signal, we subtract the input signal from the total signal.
Finally, we compute the Fourier transform of the reflected and input signals in order to compute the reflection coefficient $R = \hat{E}_r^2/\hat{E}_i^2$, where $\hat{E}_r$ and $\hat{E}_i$ are the reflected and incident signals in Fourier space. 

\begin{figure}[tb!]
  \begin{center}
    \includegraphics[width=1.0\textwidth]{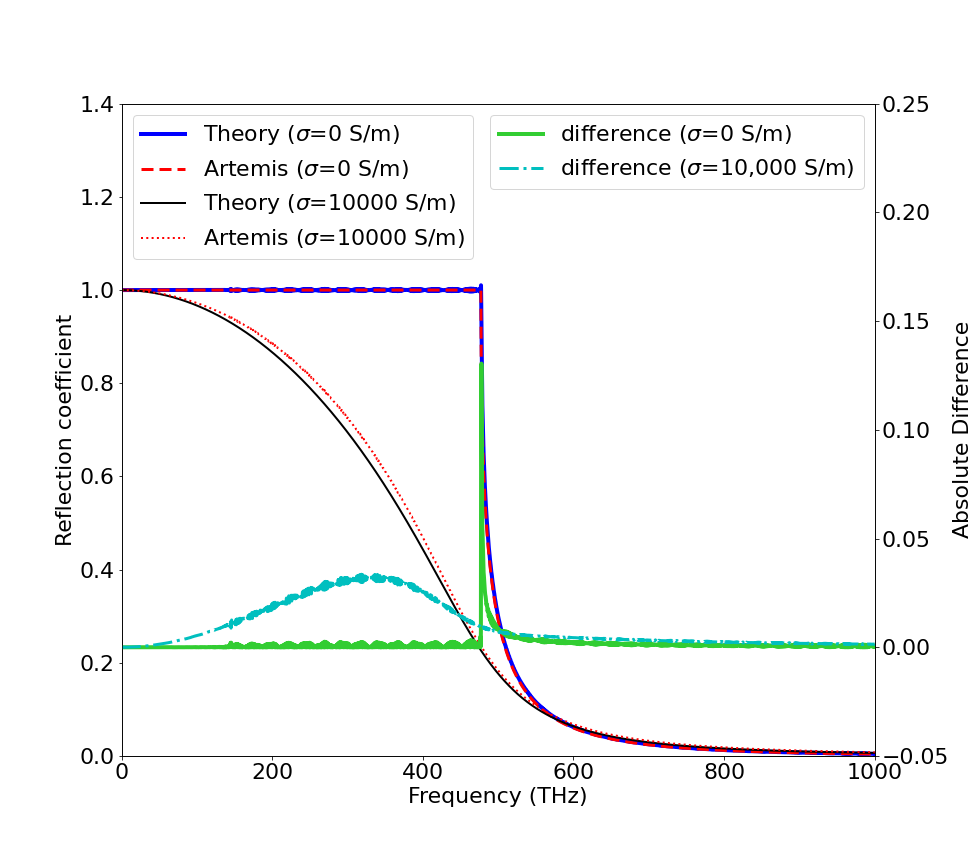}
  \end{center}
  \caption{Comparison of reflection coefficient obtained from the simulation with theory for superconductor with 100~nm London penetration depth for conductivity, $\sigma$=0 and $10^4$~S/m, \textcolor{black}{along with the absolute difference between simulation and theory.}}
    \label{fig:Refl}
\end{figure}

The comparison of the reflection coefficient obtained from our simulations with the theoretical predictions given in equation (\ref{eq:Ref}), \textcolor{black}{along with their abolute differences,} is shown in Fig.~\ref{fig:Refl}.
The results from our code match the theory \textcolor{black}{within 10\% in regions with reflection coefficient greater than 0.4, and the relative difference increases to a maximum of 30\%  as the reflection coefficient decreases for the $\sigma=0$~S/m case.}
Notable, for the case with $\sigma=0$, the cut-off frequency from both the simulation and theory are at $f_c = 1/(2\pi\lambda\sqrt{\epsilon\mu}) = 478.75$~THz.
Thus, we infer that for all operating frequencies below this frequency, the superconductor with $\sigma=0$ will behave as expected with a reflection coefficient of 1.
Also, for $\sigma=10^4$~S/m, as long as the operating frequency is below a few THz, the reflection coefficient is nearly 1 and the metal will behave in the same way as a superconductor.

\subsection{Convergence Tests}
Having validated the physical accuracy of the Maxwell-London solver, we now demonstrate the spatial and temporal convergence of our numerical implementation using three different geometrical configurations 
with superconducting materials that have finite conductivity.
\begin{itemize}
\item In the first setup \textcolor{black}{(Section \ref{sec:homogeneous})}, the three-dimensional domain is homogeneous with constant properties for the material throughout the domain and a Gaussian pulse initialized propagates through the medium. 
\item In the second setup \textcolor{black}{(Section \ref{sec:strip})}, we introduce a thin strip of material (conductive or superconductive) in a vacuum domain and study the convergence for the Gaussian pulse that interacts with the material strip.
\item In the third setup \textcolor{black}{(Section \ref{sec:box})}, to study convergence for a relatively complex \textcolor{black}{field} structure compared to the first two cases, we embed a cubic superconducting material at the center of the vacuum domain which interacts with the incident Gaussian pulse.
\end{itemize}

In each set of convergence tests, the computational domain extends from -640 to +640~nm in each of the three directions with vacuum permittivity and permeability.
We use periodic boundary conditions in all the tests described below.
\textcolor{black}{We perform tests with four types of media listed below to compare the solutions with the superconducting material with finite conductivity.}
\begin{itemize}
    \item (a) vacuum everywhere ($\sigma=0$~S/m \textcolor{black}{and superconductivity disabled}),
    \item (b) purely conductive ($\sigma=10^4$~S/m \textcolor{black}{and superconductivity disabled}),
    \item (c) purely superconducting (\textcolor{black}{$\sigma=0$ and} $\lambda=40$~nm),
    \item (d) superconducting with finite conductivity ($\sigma=10^4$~S/m and $\lambda$=40~nm).
\end{itemize}

In order to compute the convergence rates, we use a standard procedure for computing $L^1$ error norms for the $\Eb$ and $\Bb$ fields at increasing resolution in space and/or time.
This involves computing the error between ``coarse'' and ``medium'' resolution solutions, $E_{\rm coarse}^{\rm medium}$, and the error between ``medium'' and ``fine'' resolution solutions, $E_{\rm medium}^{\rm fine}$.
The exact procedure, which includes the spatial averaging procedure for the $\Eb$ and $\Bb$ fields at different resolutions, is described in detail in Section 5.1 of the original ARTEMIS paper \cite{yao2022massively}.
We note that for the temporal-only convergence tests below, no spatial averaging is required since all solutions use the same cell size.

\subsubsection{Homogeneous Conductive and Superconducting Domains}
\label{sec:homogeneous}
\textcolor{black}{
First, we demonstrate that the algorithm within homogeneous domains is second-order in space and time with all the physics turned on, i.e., with the superconducting material with finite conductivity. We perform four tests where the entire domain is homogeneous with the material types listed in cases (a)-(d) above.}
We initialize three Gaussian pulses, given by,
\begin{eqnarray}
E_x = e^{-z^2/L^2}, &\quad& B_x = \frac{1}{c}e^{-y^2/L^2}, \nonumber\\
E_y = e^{-x^2/L^2}, &\quad& B_y = \frac{1}{c}e^{-z^2/L^2}, \nonumber\\
E_z = e^{-y^2/L^2}, &\quad& B_z = \frac{1}{c}e^{-x^2/L^2},
\end{eqnarray}
with $L=80$~nm and $c=1/\sqrt{\epsilon_0\mu_0}$~m/s.
Note that these fields are consistent with the relationship satisfying the intrinsic impedance in vacuum, and as demonstrated below, result in pure translation under vacuum conditions.
In each simulation, the domain is discretized with $512^3$ grid cells (2.5~nm resolution) and the simulation is performed to 200 time steps with a CFL of 0.9 to a physical time of 0.87~fs.
The three Gaussian pulses are all initialized to propagate in different directions, so showing one-dimensional plots of any one field is rotationally equivalent to the results for the other fields.

\begin{figure}[tb!]
    \centering
    \includegraphics[width=\linewidth]{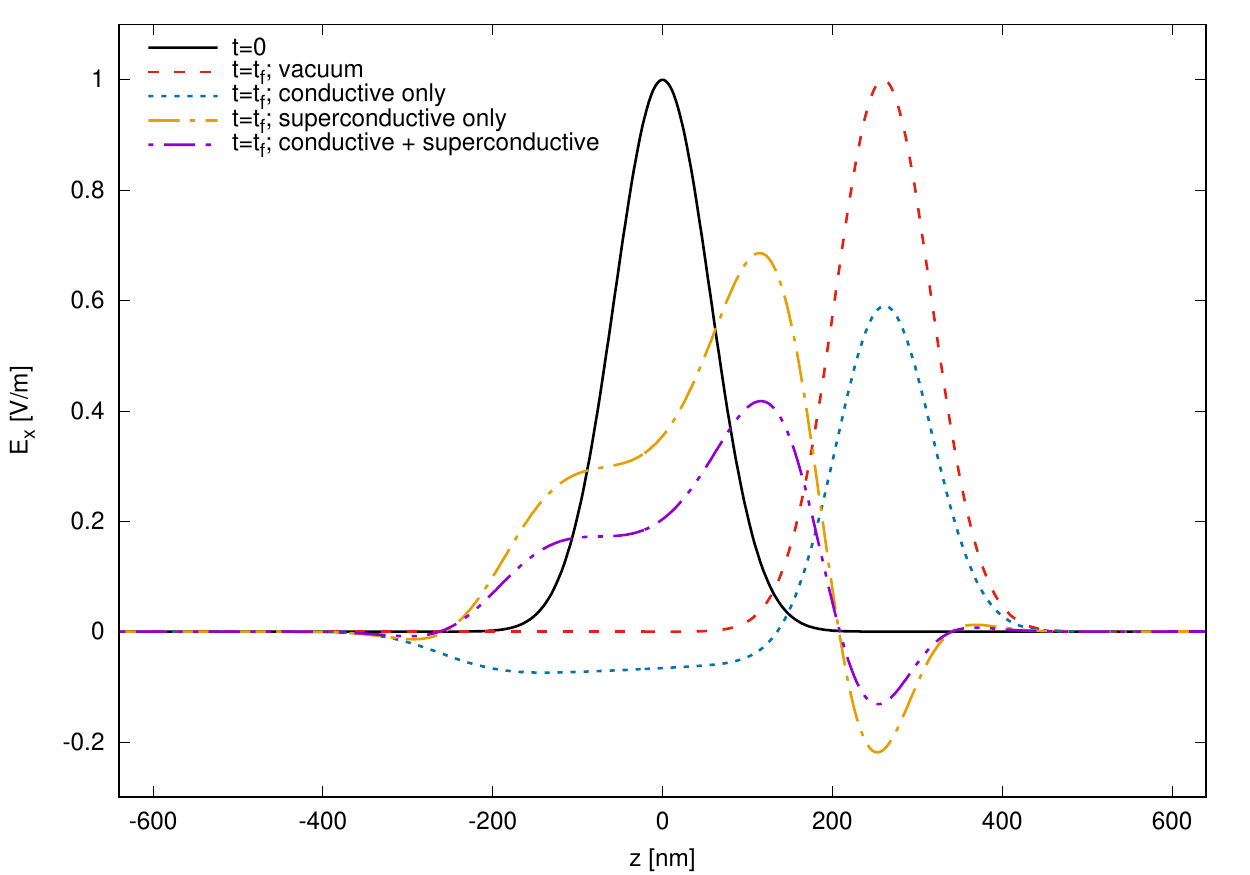}
    \caption{Initial and final $E_x$ field as a function of $z$ with homogeneous medium with the domain. We compare the variation of $E_x$ for the cases where the domain is vacuum everywhere, conductive only ($\sigma=10^4$~S/m), superconducting with finite conductivity ($\sigma=10^4$~S/m and $\lambda$=40~nm).}
    \label{fig:pulse_all}
\end{figure}

First, to demonstrate that both the conductive physics and superconductive physics have nontrivial contributions to the evolution of the system, we compare the results obtained from the four simulations with material types listed above in (a)-(d).
In Fig.~\ref{fig:pulse_all}, we show the initial and final configuration of $E_x(z)$, extracted along the $z-$direction at the center of the domain, to show the nontrivial contribution of physics for each type of material.
Case (a) with homogeneous vacuum domain results in pure translation as expected.
For Case (b) with finite conductivity, the signal undergoes attenuation along with some dispersion resulting in negative values for the electric field. \textcolor{black}{This observation is a natural consequence of the frequency-dependent oscillation of the electromagnetic field.}
The signal in Case (c) undergoes some translation, and exhibits a more pronounced dispersion due to superconductivity and the complex wavenumber previously described in Section \ref{sec:skindepth}.
Finally, the signal in Case (d) is similar to Case (c) with additional attenuation due to the finite conductivity.

Next, we compute the numerical convergence in space and time (simultaneously) for Case (d) where both the conductive and superconductive physics is enabled in the entire domain. 
To compute the convergence, we perform 3 simulations using a computational mesh with $128^3, 256^3$, and $512^3$ grid cells (i.e., 10, 5, and 2.5~nm cell resolution, respectively).
We use a CFL of 0.9 for each simulation and run the simulations for 50, 100, and 200 time steps, i.e., to the same physical time of 0.87~fs.
In Table \ref{tab:pure_sc_spacetime_convergence} we show clear second-order convergence in space and time for all field components.
\begin{table}[tb!]
\centering
\begin{tabular}{p{2.5cm} p{3cm} p{3cm} p{2cm}}
\hline
Variable & $E_{\rm coarse}^{\rm medium}$ & $E_{\rm medium}^{\rm fine}$ & Rate\\
\hline
$E_x, E_y, E_z$ & $3.33\times 10^{-4}$ & $8.30\times 10^{-5}$ & 2.00 \\
$B_x, B_y, B_z$ & $6.05\times 10^{-13}$ & $1.52\times 10^{-13}$ & 1.99 \\
\hline
\end{tabular}
\caption{Spatial and temporal convergence rates for a Gaussian pulse propagating through a homogeneous superconducting domain with finite conductivity ($\lambda=40$~nm and $\sigma=10^4$~S/m).}
\label{tab:pure_sc_spacetime_convergence}
\end{table}

\subsubsection{Conductive and Superconductive Strips}
\label{sec:strip}
\textcolor{black}{In these tests, the domain is vacuum except for a thin strip (in $z$).} 
Specifically, a thin strip of material is initialized from $z=-40$~nm to $z=+40$~nm, and it extends to the periodic domain boundaries in $x$ and $y$. Similar to the homogeneous domain cases presented above, to demonstrate that both conductive and superconductive physics have nontrivial contributions to the evolution of the system, we perform the 4 simulations with the same parameters for $\sigma$ and $\lambda$ described in the previous section (Section \ref{sec:homogeneous}). 
We initialize a single Gaussian pulse given by,
\begin{eqnarray}
E_x = e^{-(z-z_0)^2/L^2}, &\quad& B_x = 0, \nonumber\\
E_y = 0, &\quad& B_y = \frac{1}{c}e^{-(z-z_0)^2/L^2}, \nonumber\\
E_z = 0, &\quad& B_z = 0,\label{eq:one_plane}
\end{eqnarray}
with $z_0=-320$~nm.
Even though we perform the simulations in three-dimensions, this setup is essentially one-dimensional since the wave propagates purely in the $z$ direction, and $E_y, E_z, B_x$, and $B_z$ remain zero. The simulation domain is discretized with a $512^3$ grid (i.e., 2.5~nm resolution) and is performed for 400 timesteps with CFL$=$0.9 to a physical time of 1.73~fs.

\begin{figure}[tb!]
    \centering
    \includegraphics[width=\linewidth]{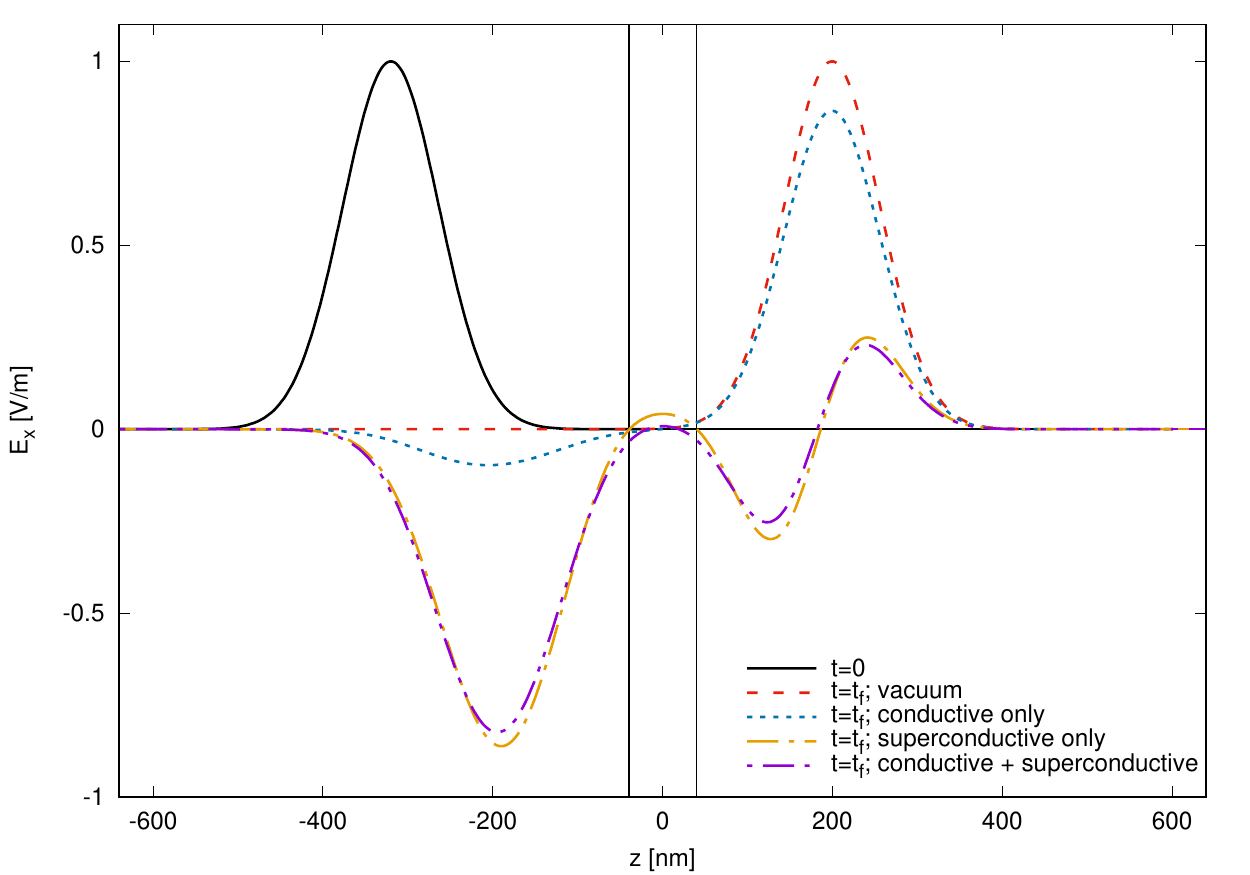}
        \caption{Initial and final $E_x$ field as a function of $z$ for a Gaussian pulse initialized in vacuum interacting with a strip of material indicated by the region between vertical solid lines. We compare the variation of $E_x$ for the cases where the domain is vacuum everywhere, conductive only ($\sigma=10^4$~S/m), superconducting with finite conductivity ($\sigma=10^4$~S/m and $\lambda$=40~nm).}
    \label{fig:pulse_flat}
\end{figure}

In Fig.~\ref{fig:pulse_flat} we show the initial and final configuration of $E_x(z)$, extracted along the $z$-direction at the center of the domain, to show the nontrivial evolution of the signal after it interacts with the strip, indicated by vertical lines.
It can be seen that, similar to the homogeneous setup, Case (a) with pure vacuum condition results in pure translation of the signal.
For Case (b), the signal interacts with the conductive strip and undergoes attenuation as well as reflection evident from the negative value of the signal upstream of the metal strip.
For the superconducting strip in Case (c), we see a more complex signal in the final step of the simulation where the pulse is modified by two superconducting (zero conductivity) interfaces. After interacting with the material strip, the pulse undergoes reflection, and transmission with frequency-dependent dispersion.
Finally, in Case (d) with superconducting strip with finite conductivity, the final profile of the signal is similar to Case (c) but with additional attenuation. 

\begin{table}[tb!]
\centering
\begin{tabular}{p{2.5cm} p{3cm} p{3cm} p{2cm}}
\hline
Variable & $E_{\rm coarse}^{\rm medium}$ & $E_{\rm medium}^{\rm fine}$ & Rate \\
\hline
$E_x$ & $1.23\times 10^{-4}$ & $3.08\times 10^{-5}$ & 2.00 \\
$B_y$ & $3.90\times 10^{-13}$ & $9.73\times 10^{-14}$ & 2.00 \\
\hline
\end{tabular}
\caption{Temporal convergence rates for a Gaussian pulse interacting with a strip of superconducting material with finite conductivity ($\lambda=40$~nm and $\sigma=10^4$~S/m).}
\label{tab:flat_time_convergence}
\end{table}
Since this configuration includes an interface, we perform separate tests to compute temporal and spatial convergence.
We first perform tests with Case (d) to demonstrate that the algorithm is second-order in time in the presence of both superconducting and conductive currents. 
We perform 3 simulations with a computational mesh containing $256^3$ grid cells (i.e., 5~nm cell resolution), but use a CFL of 0.9, 0.45, and 0.225 to vary the timestep resolution, such that the simulations reach the same physical time of 1.73~fs using 200, 400, and 800 time steps, respectively.
In Table \ref{tab:flat_time_convergence} we show clear second-order convergence in time for all the field components.

\begin{table}[tb!]
\centering
\begin{tabular}{p{2.5cm} p{3cm} p{3cm} p{2cm}}
\hline
Variable & $E_{\rm coarse}^{\rm medium}$ & $E_{\rm medium}^{\rm fine}$ & Rate \\
\hline
$E_x$ & $9.05\times 10^{-3}$ & $4.86\times 10^{-3}$ & 0.90 \\
$B_y$ & $2.82\times 10^{-11}$ & $1.53\times 10^{-11}$ & 0.88 \\
\hline
\end{tabular}
\caption{Spatial convergence rates for a Gaussian pulse interacting with a strip of superconducting material with finite conductivity ($\lambda=40$~nm and $\sigma=10^4$~S/m).}
\label{tab:flat_space_convergence}
\end{table}
Next, we conduct a spatial-only convergence test for Case (d) to demonstrate that the algorithm is first-order in space, even in the presence of the spatial discontinuity in material properties (i.e., vacuum-superconducting interfaces). 
We perform 3 simulations with $128^3, 256^3$, and $512^3$ grid cells (i.e., 10, 5, and 2.5~nm cell resolution, and using a CFL of 0.225, 0.45, and 0.9 for each simulation, such that the timestep in each simulation is the same.
We run all the simulations for 400 time steps (to the same physical time of 1.73~fs).
Due to the abrupt spatial discontinuity in both physics and conductivity, we see in Table \ref{tab:flat_space_convergence} that the algorithm is first-order in space, which is expected for configurations with inherent discontinuities.
\textcolor{black}{We have confirmed with separate simulations that for the case where $\sigma=0$ and $\lambda$ smoothly varies from 400~nm to 40~nm over the entire right-half of the domain that second-order spatial convergence is retained.  In other words, this is the case where $1/\lambda$ transitions relatively smoothly at the material interface and we expect second-order convergence.}

\subsubsection{Cubic Block of Superconductive Material}
\label{sec:box}
\begin{figure}[tb!]
    \centering
    \includegraphics[width=0.43\linewidth]{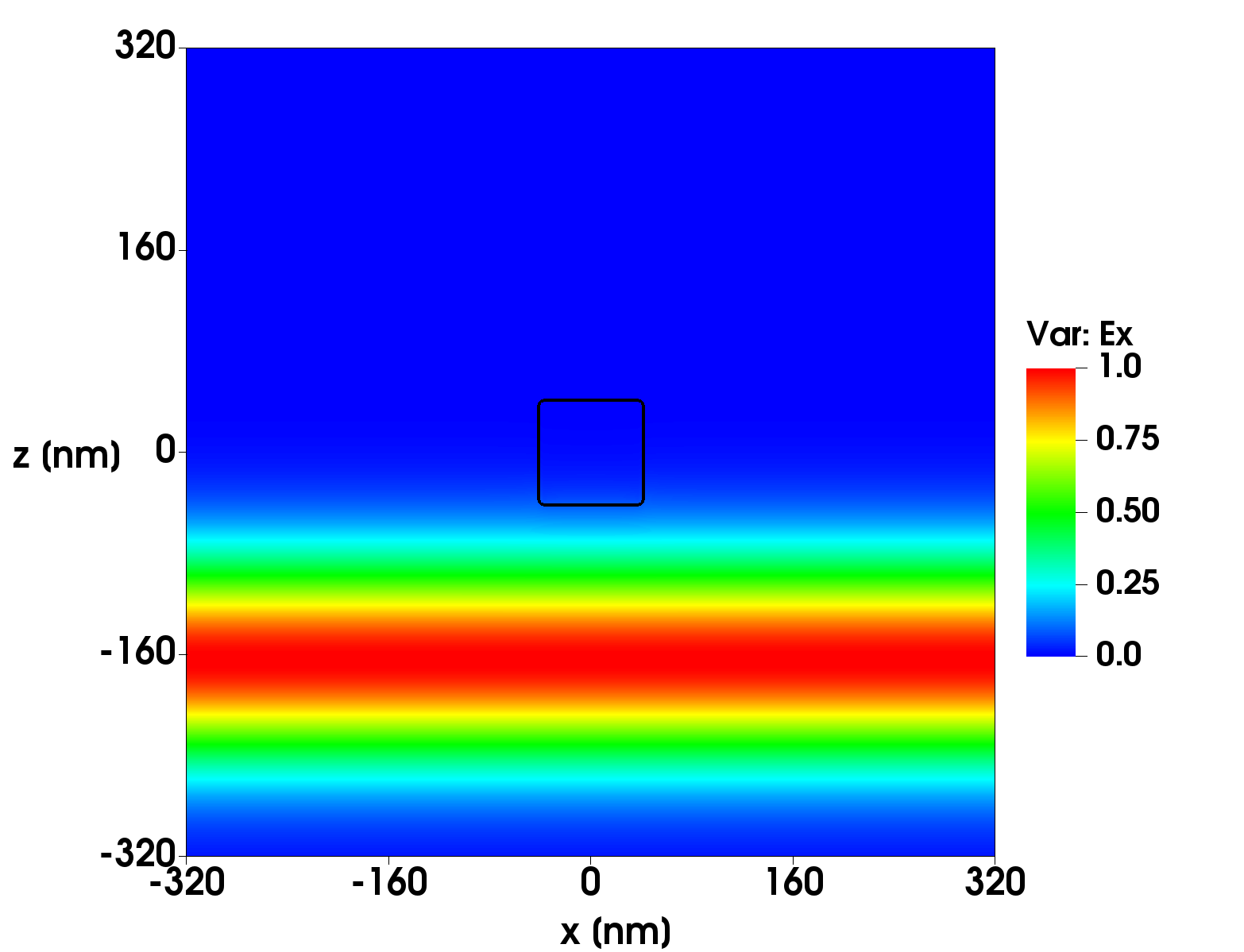}
    \includegraphics[width=0.43\linewidth]{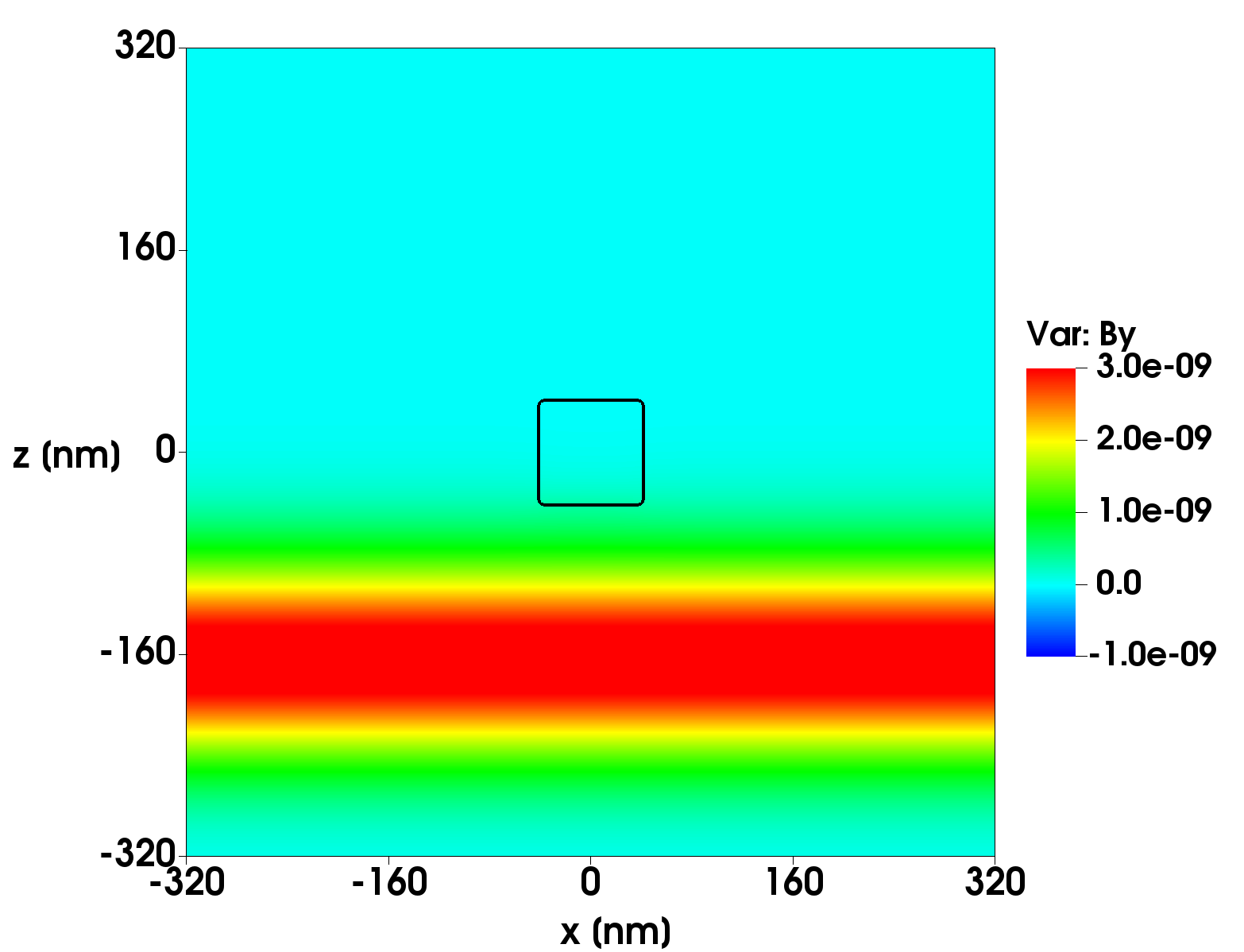}\\
    \includegraphics[width=0.43\linewidth]{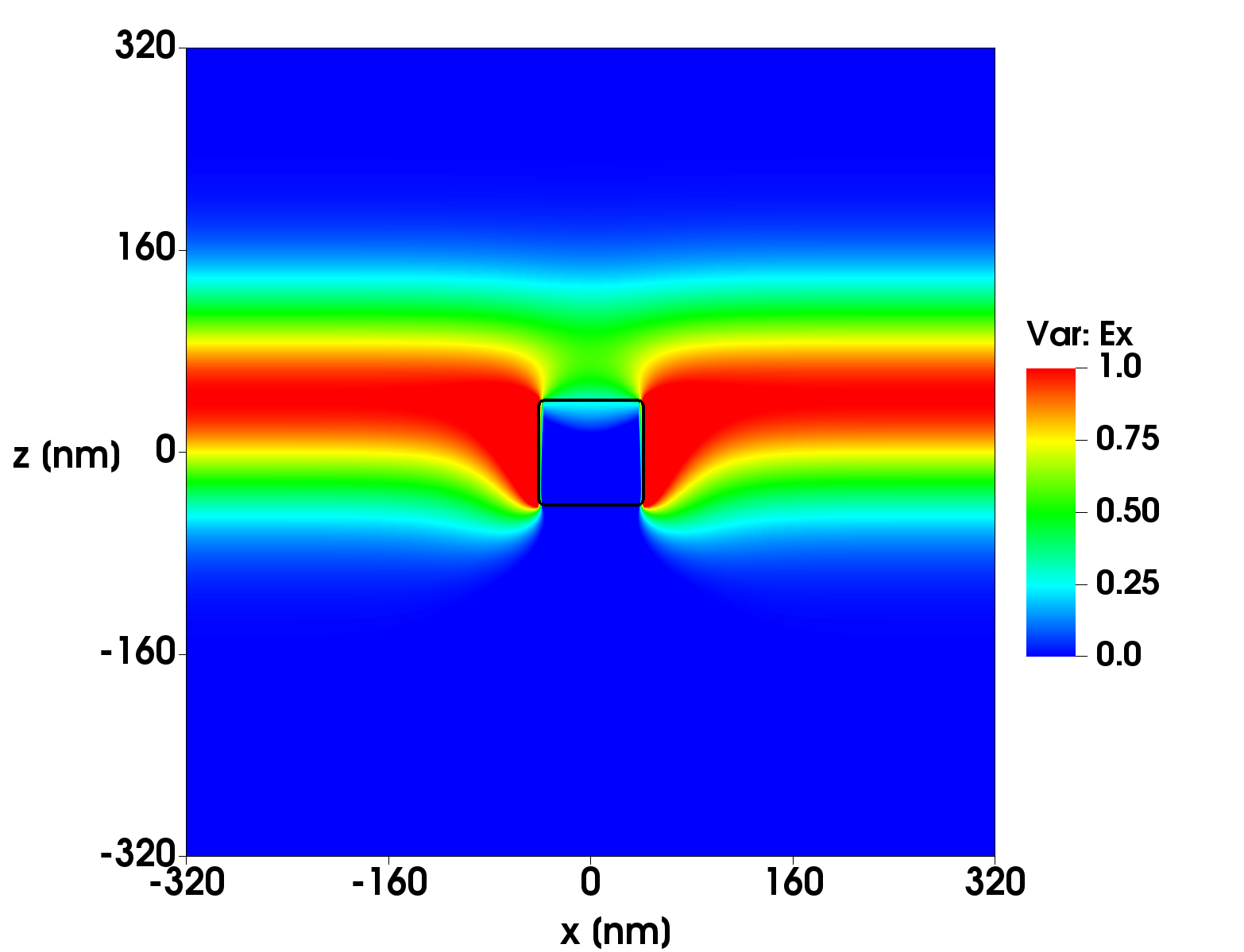}
    \includegraphics[width=0.43\linewidth]{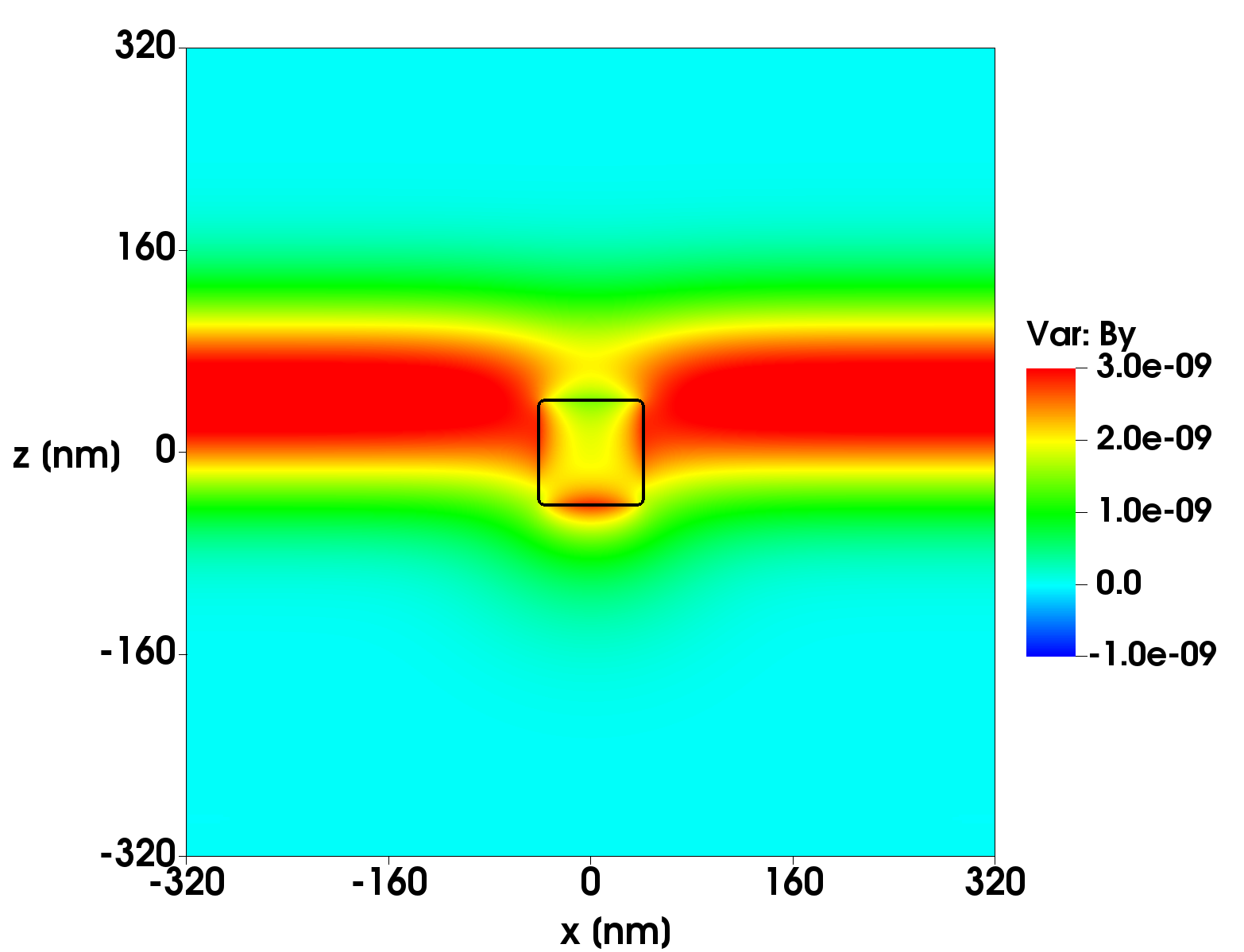}\\
    \includegraphics[width=0.43\linewidth]{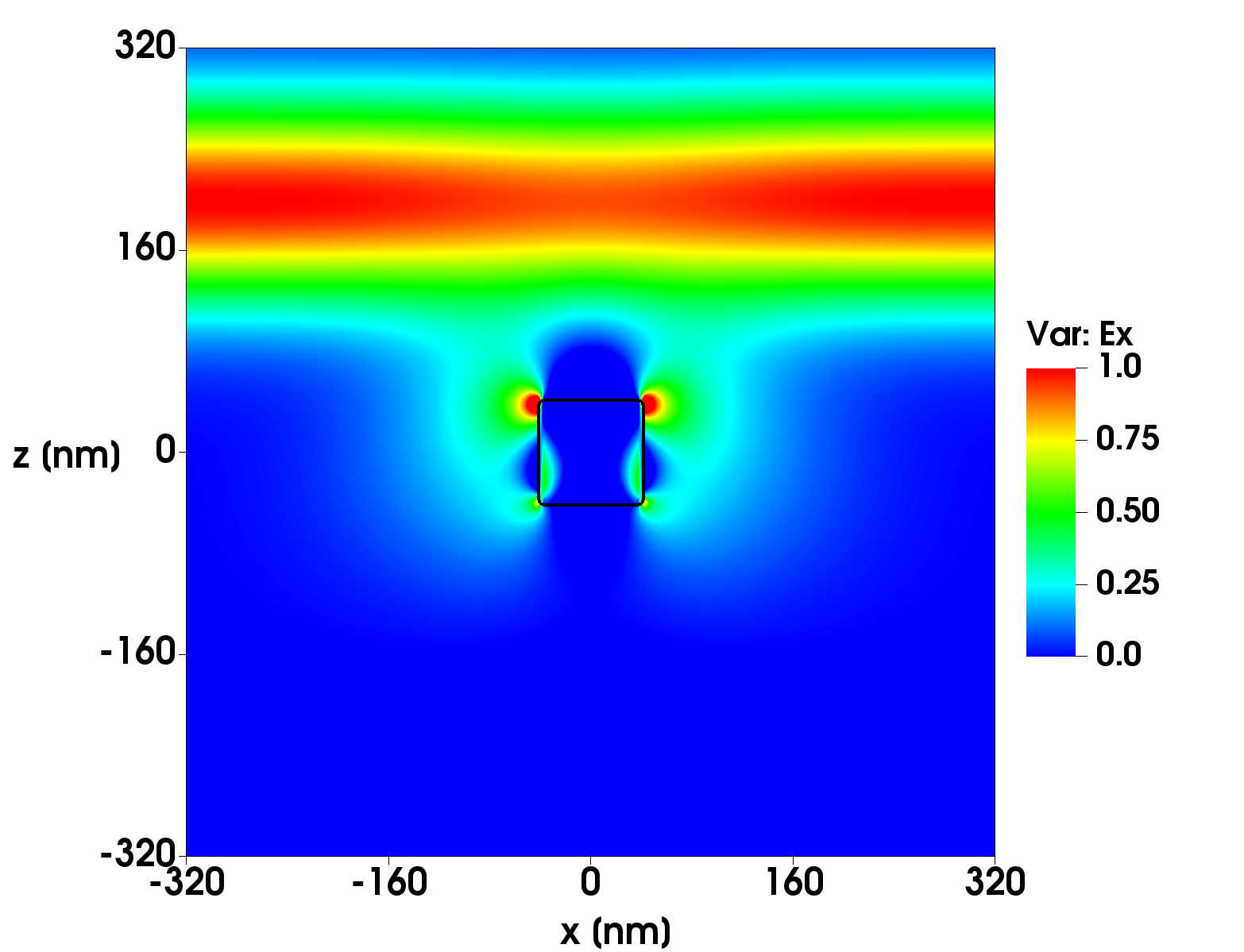}
    \includegraphics[width=0.43\linewidth]{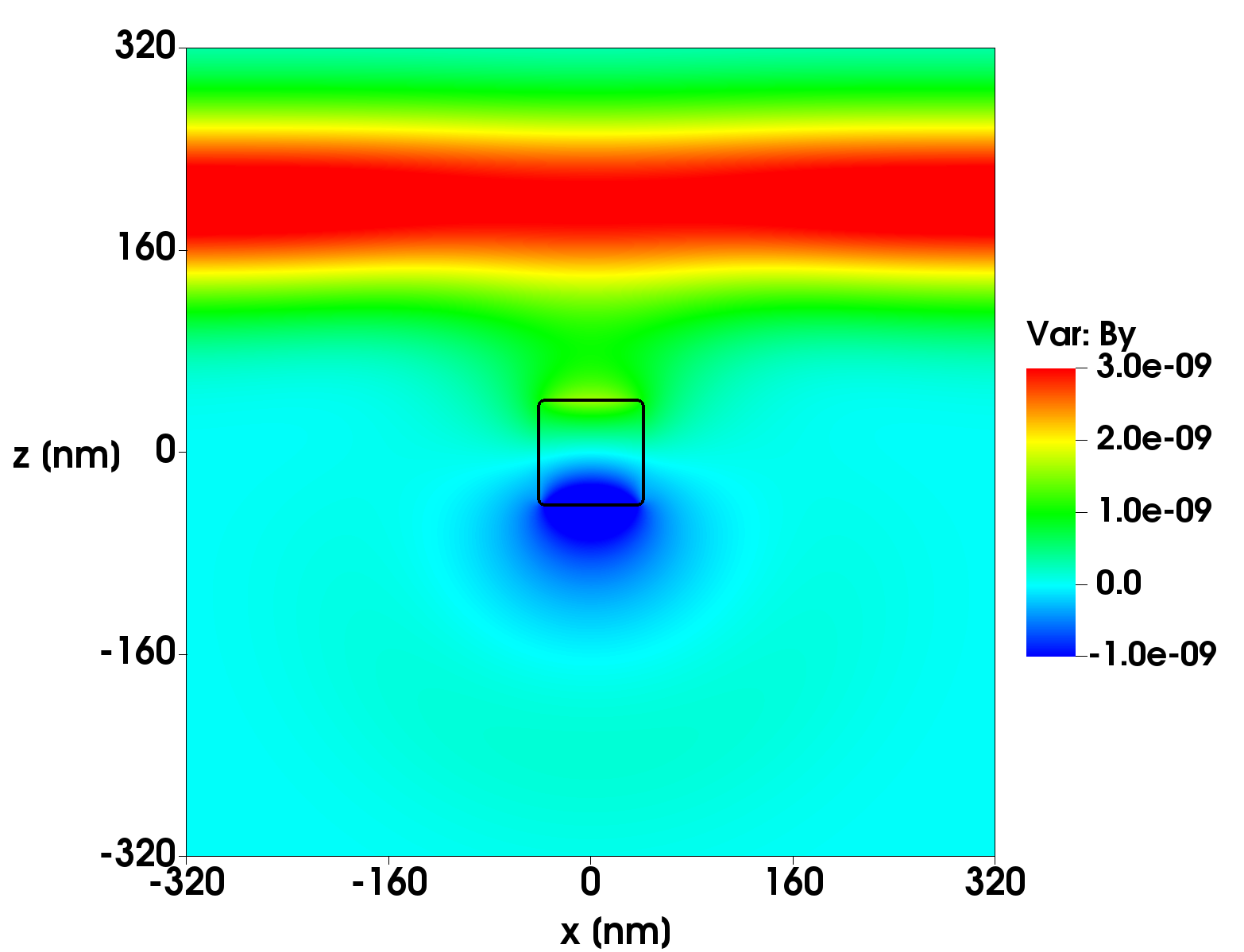}
    \caption{
    Time-evolution of field components, $E_x$ (left column) and $B_y$ (right column) extracted along an $x-z$ slice at the center of the three-dimensional domain at $t=0.52$ (top,) 1.21 (middle), and 1.73 (bottom)~fs. Figures show a zoomed-in view of the slice, with the Gaussian pulsar propagating along the $z$-direction through an embedded cube of superconducting material with finite conductivity ($\sigma=10^4$~S/m and $\lambda=40$~nm indicated as the solid black box)}
    \label{fig:sc_cube_time}
\end{figure}

\textcolor{black}{In these set of tests, a cubic material with material type (d) is embedded in a vacuum domain to demonstrate the spatial and temporal convergence for more complex three-dimensional field structures compared to the first two configurations used for convergence studies.}
The setup and initialization for this case is identical to the previous case with a material strip, except that here the domain has an embedded cube extending from $-40$~nm to $+40$~nm in all three spatial directions. 
For this set up, we only consider Case (d) where the material is superconducting with finite conductivity ($\lambda=40$~nm and $\sigma=10^4$~S/m).
Even though we initialize only $E_x$ and $B_y$ components, (described previously in Section \ref{sec:strip}), as the Gaussian pulse propagates along the $z-$direction, it interacts with the cubic block of material and all components of the electric and magnetic field develop complex structures \textcolor{black}{as seen in Fig.~\ref{fig:sc_cube_time}}, thus allowing us to study convergence in three-dimensions for all components.

\begin{table}[tb!]
\centering
\begin{tabular}{p{2.5cm} p{3cm} p{3cm} p{2cm}}
\hline
Variable & $E_{\rm coarse}^{\rm medium}$ & $E_{\rm medium}^{\rm fine}$ & Rate \\
\hline
$E_x$ & $1.02\times 10^{-4}$ & $2.55\times 10^{-5}$ & 2.00 \\
$E_y$ & $7.38\times 10^{-7}$ & $1.84\times 10^{-7}$ & 2.00 \\
$E_z$ & $7.67\times 10^{-7}$ & $1.91\times 10^{-7}$ & 2.00 \\
$B_x$ & $1.49\times 10^{-16}$ & $3.72\times 10^{-17}$ & 2.00 \\
$B_y$ & $3.40\times 10^{-13}$ & $8.49\times 10^{-14}$ & 2.00 \\
$B_z$ & $4.71\times 10^{-15}$ & $1.18\times 10^{-15}$ & 2.00 \\
\hline
\end{tabular}
\caption{Temporal convergence rate for a Gaussian pulse propagating through a cube of superconducting material with finite conductivity ($\lambda=40$~nm and $\sigma=10^4$~S/m).}
\label{tab:cube_time_convergence}
\end{table}

\begin{table}[tb!]
\centering
\begin{tabular}{p{2.5cm} p{3cm} p{3cm} p{2cm}}
\hline
Variable & $E_{\rm coarse}^{\rm medium}$ & $E_{\rm medium}^{\rm fine}$ & Rate \\
\hline
$E_x$ & $1.68\times 10^{-3}$ & $4.77\times 10^{-4}$ & 1.82 \\
$E_y$ & $1.31\times 10^{-4}$ & $6.48\times 10^{-5}$ & 1.01 \\
$E_z$ & $1.44\times 10^{-4}$ & $7.18\times 10^{-5}$ & 1.00 \\
$B_x$ & $7.89\times 10^{-14}$ & $3.83\times 10^{-14}$ & 1.04 \\
$B_y$ & $5.37\times 10^{-12}$ & $1.46\times 10^{-12}$ & 1.88 \\
$B_z$ & $5.54\times 10^{-13}$ & $2.81\times 10^{-13}$ & 0.98 \\
\hline
\end{tabular}
\caption{Spatial convergence rate for a Gaussian pulse propagating through a cube of superconducting material with finite conductivity ($\lambda=40$~nm and $\sigma=10^4$~S/m).}
\label{tab:cube_space_convergence}
\end{table}
In Fig.~\ref{fig:sc_cube_time}, we show the time evolution of only two components, $E_x$, and $B_y$, extracted on an $x-z$ slice through the center of the domain, with the embedded box at the center. 
The figure illustrates that both $E_x$ and $B_y$ fields develop complex structures, especially near the embedded box, as the signal propagates through the superconductor from $z<0$ to $z>0$.
The figures at the top show the incident Gaussian pulse at 0.52~fs on the embedded box. As the pulse propagates through the center of the domain, we observe that the signal is mainly transmitted outside the embedded box (evident from signal at 1.21~fs), while, inside the box, the amplitude of the $E_x$ and $B_y$ components is very small.
Finally, as the signal completely propagates through the material, we observe that surface fields develop surrounding the embedded box, and we attribute this mainly to the superconducting current that may evolve in this regions. 
We would like to note that, the main purpose of this setup is to perform numerical convergence tests in three-dimensions.
We conduct separate tests for temporal and spatial convergence, similar to the tests performed for the material strip in Section \ref{sec:strip}, and obtain the same overall conclusions, namely, second-order accuracy in time and first-order accuracy in space, as illustrated in Tables \ref{tab:cube_time_convergence} and \ref{tab:cube_space_convergence}, respectively.

\section{Coplanar Waveguide Resonator}\label{sec:coplanar}

\begin{figure}[tb!]
    \centering
    \includegraphics[width=0.96\linewidth]{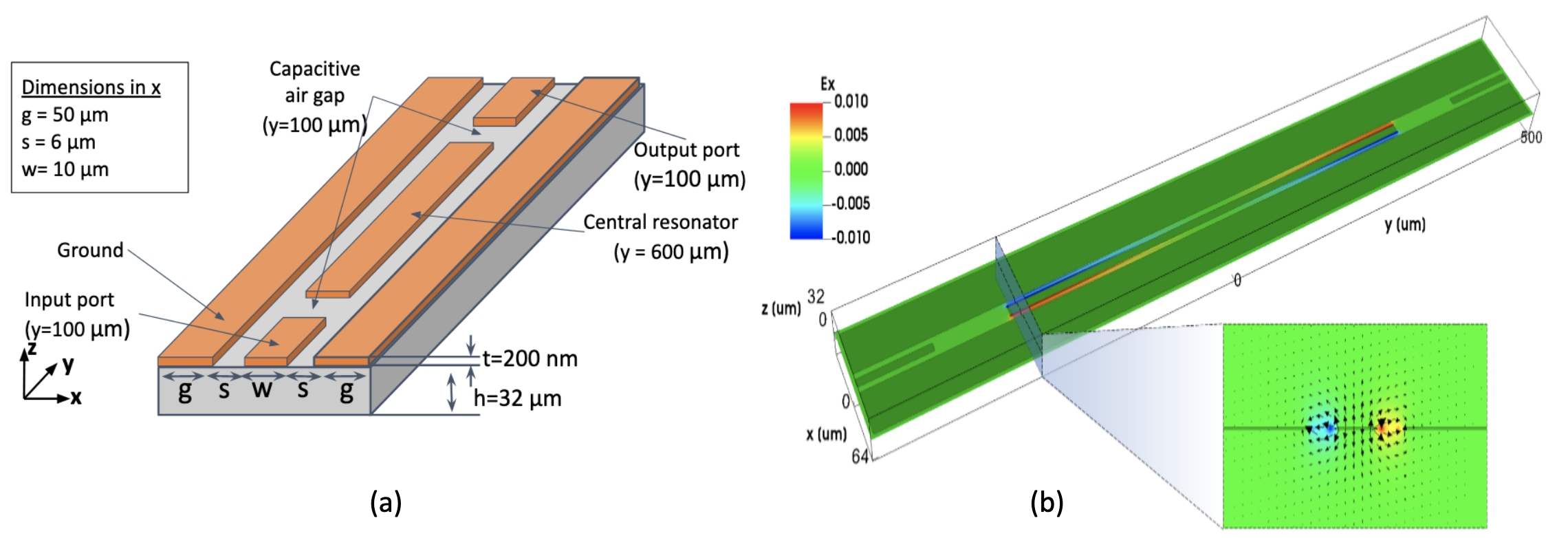}
    \caption{(a) Schematic to illustrate a CPW resonator structure found in quantum readout applications with superconducting films sitting atop a silicon substrate. 
              (b) Spatial variation of electric field along an $x-z$ slice passing through the transmission lines. The dark shaded regions indicate metal (either conducting or superconducting).  The red and blue shading indicates the magnitude of the $E_x$ field (blue/red = $\pm 0.001$~V/m) near the end of the simulation, illustrating the fundamental mode.  The inset is an $x-z$ slice with normal in the $y$-direction extract at the front of the resonator line, $y=-300~\mu$m, with vectors illustrating the electric field.}
    \label{fig:CPW_ab}
\end{figure}

In this section, we present three-dimensional simulations performed using ARTEMIS for a coplanar waveguide (CPW) resonator verifying that we capture the resonant behavior of the structure and we also present the $Q$-factor measurements for different material configurations. In our simulations, the computational domain has a physical size of 
[-65,65]~$\mu$m in $x$,
[-504,504]~$\mu$m in $y$, and
[0,64]~$\mu$m in $z$.
\textcolor{black}{The domain size in $z$ was chosen to be large enough to prevent loss of information of the closed loop magnetic field lines by verifying matching results with larger domain sizes in $z$ over shorter time (to save computational resources).}
We discretize the domain with $130\times 1008\times 1280$ grid cells, so that $\Delta x = \Delta y = 1~\mu$m and $\Delta z = 50$~nm.
In Figure~\ref{fig:CPW_ab}(a), we show a schematic of the CPW resonator to illustrate the resonator structure used in our simulations. We use a CPW structure designed to support a fundamental mode of approximately $100$~GHz based on approximate analytic formulas \cite{chen1997characteristics}.
Note that, our circuit dimensions and design frequency are based on typical length scales and operating conditions used in quantum readout applications \cite{sage2011study}.

The resonator structure, as shown in Fig.~\ref{fig:CPW_ab}(a) consists of a silicon substrate, with a thickness of $h=32~\mu$m and relative permittivity of $11.7$.  Thin superconducting films sit atop the silicon substrate with a thickness of $t=200$~nm and relative permittivity of 1. The remainder of the domain on top of the resonator structure is vacuum.  The grid cell size in the $z$-dimension, $\Delta z = 50$~nm, is chosen to properly resolve the finest-scale feature in the simulation, which in this case is the superconducting film with $200$~nm thickness. The central resonator line has a length of $600~\mu$m along the $y-$direction, the input/output ports have a length of $100~\mu$m and the air gaps between the central resonator line and the input/output ports is $100~\mu$m in $y$, allowing for capacitive coupling between the ports and resonator. All three components have a width of $w=10~\mu$m. The ground planes have a length of $1000~\mu$m in $y$ aand width $g=50~\mu$m, and the air gap between the ground planes and the input/output as well as transmission lines is $s=6~\mu$m.
The relative permeability everywhere in the domain is 1. We use a perfectly matched layer (PML) \cite{berenger1994perfectly,berenger1996three} boundary condition on all domain faces.
Note that with these geometrical specifications of the CPW, we use a $4~\mu$m vacuum gap between the domain boundary and the outer $x$ and $y$ edges of the CPW resonator\textcolor{black}{; thus each PML boundary condition is in contact with either vacuum or dielectric material to allow for signals to propagate out of all domain faces. The use of PML in contact with superconducting material in the two-fluid model is not well-understood and is a subject for future work.}

In the air gaps between the ground planes and the input port, we provide two (opposite in magnitude in the left and right air-gaps) soft-source excitations.
The excitation used in the two air-gaps at the front end, i.e., at $y=-500~\mu$m is a modulated sine wave with center frequency, $f_{\rm in}$, and associated period $T_{\rm in} = 1/f_{\rm in}$. 
\begin{equation}
    E_x = \pm e^{-(t-4T_{\rm in})^2 / (2T_{\rm in}^2)}\sin{\left(\frac{2\pi t}{T_{\rm in}}\right)}~\textrm{V/m},
\end{equation}
In each simulation we choose $f_{\rm in} = 100$~GHz to match the predicted resonance frequency of the CPW. 
We use a CFL of 0.95, corresponding to a time step of $\Delta t \approx 0.158$~fs.
We run our simulations on 32 NVIDIA A100 GPUs on the NERSC perlmutter system for 12 hours, to a total time of $\sim 134$~ps ($\sim 850,000$ time steps, and we find that each time step requires $\sim 0.05$~s).
While we have resolved the superconducting film thickness in $z$, ideally, we would also like to resolve the structure in $x$ and $y$, and also study the effects of varying circuit dimensions.
However, computational allocations limit us to this size system and a limited number of simulations with a high-resolution only along the $z-$direction, which may be sufficient for the analysis we present.
We conduct four tests to measure the effects of using the two-fluid model for superconductivity when compared to traditional approximations for superconductivity, such as, modeling the material as a regular or perfect conductor with artificially high conductivity.
Below are the material properties we set for the superconducting material in each of the four test cases:
\begin{itemize}
    \item Case 1: Regular conductor ($\sigma=6\times 10^7$ S/m \textcolor{black}{and superconductivity disabled})
    \item Case 2: Regular conductor, but with artificially high conductivity ($\sigma=10^{10}$ S/m \textcolor{black}{and superconductivity disabled}),
    \item Case 3: \textcolor{black}{Superconductor with finite conductivity ($\sigma=6\times 10^7$ S/m and $\lambda=100$~nm),}
    \item Case 4: \textcolor{black}{Purely superconductive ($\sigma=0$ and $\lambda=100$~nm).}
\end{itemize}
Case 1 represents a standard conductor, in this case copper, and requires only the Maxwell solver.  Case 2 represents a commonly-used high conductivity approximation for superconducting behavior, requiring only the Maxwell solver.
Case 3 represents a superconductor that has been cooled to just below the critical threshold and retains its standard conductivity properties (we keep the conductivity of copper here, to compare with Case~1, even though a typical superconductor such as niobium has a room temperature conductivity that is an order of magnitude smaller), and in this case, the two-fluid model implemented in ARTEMIS is used. Finally, Case 4 represents a superconductor that has been cooled to near absolute zero, and has essentially no conductive current, and is simulated using the two-fluid model.

In Figure~\ref{fig:CPW_ab}(b), we show the spatial variation of the $E_x$ field component along the $x$-$y$ slice such that it cuts through the thin-film to include the air-gap (transmission line) between the ground plane and input/output port and the central resonator line.
The $E_x$ field shown in the figure is obtained from the results of the Case 3 simulation at time, $t\sim 0.13$~ns. 
We observe that the system excited with the signal near the input port, achieved resonance with maximum field amplitude near the front and back edges of the transmission lines (i.e. in the air-gap region between the ground plane and the central resonator line).
As expected, the fundamental mode ($\lambda_{EM}/2$, where $\lambda_{EM}$ is the effective wavelength) of EM resonance is excited at 100 GHz. We also show a zoomed-in view of field extracted along an $x$-$z$ slice near the front edge of the central resonator (at $y=-300~\mu$m) along with the electric field vectors to demonstrate that the left and right air-gaps have opposite $E_x$ fields, but their amplitude is maximum in this region.

\begin{figure}[tb!]
    \centering
    \includegraphics[width=1\linewidth]{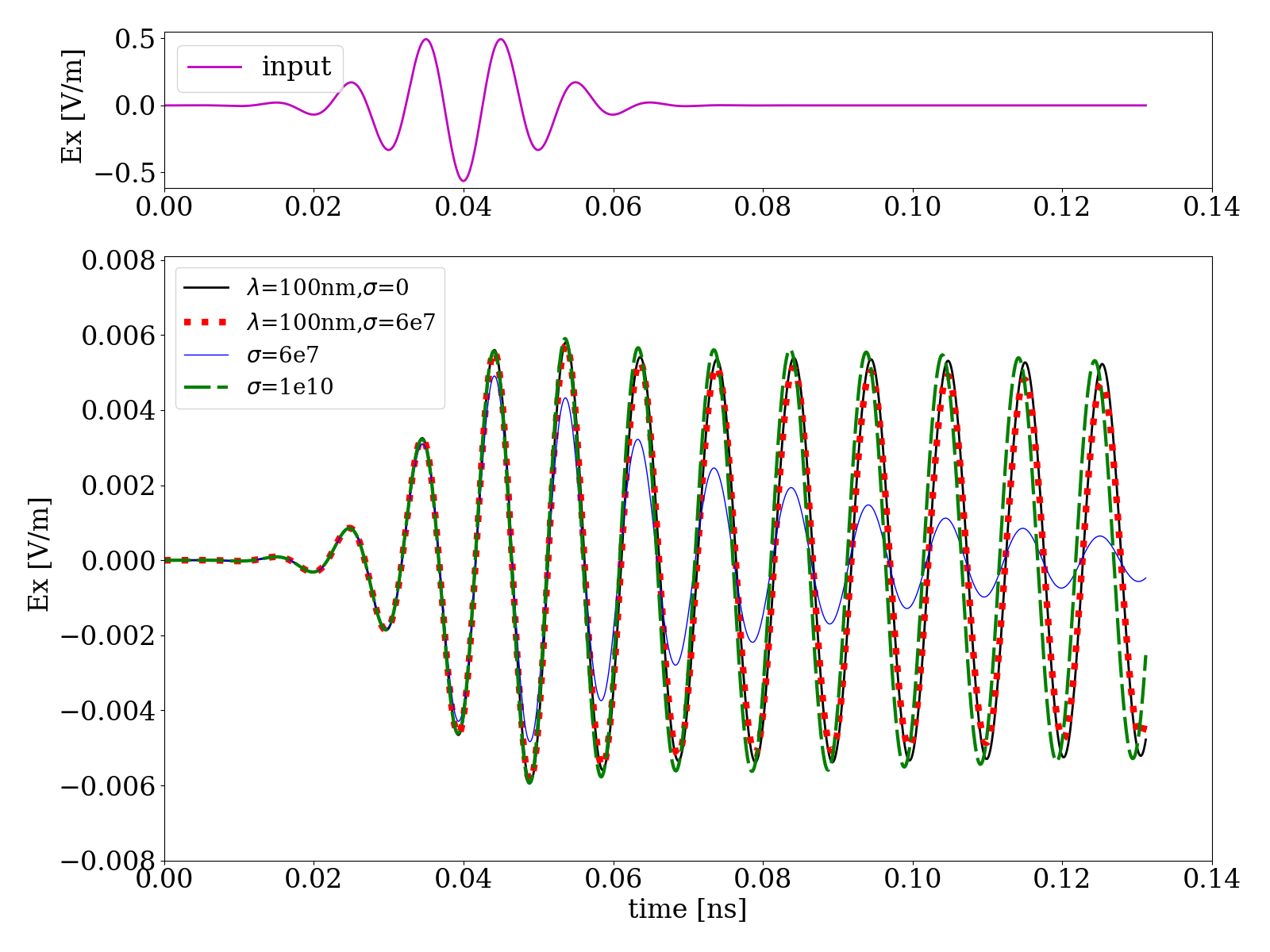}
    \caption{(Top) Measured $E_x$ field in the air gap between the input port and the ground plane, halfway down the input port.  (Bottom) Measured $E_x$ field for the four simulations in the air gap between the resonator line and the ground plane at the front of the resonator line.}
    \label{fig:CPW_plot}
\end{figure}

To further visualize resonance and compare the signal evolution among the four tests cases described above, we measure the signal, i.e., the $E_x$ field component at two locations;
(1) in the air-gap between the ground plane and input port, halfway along the length of the port, at $y=-450\mu$m to obtain the input excitation; 
(2) in the air-gap between the front edge of the resonator line and the ground plane, i.e., $y=-300~\mu$m, the same location where we observe maximum field amplitude in the $x-z$ inset in Fig.~\ref{fig:CPW_ab}(b).
In Fig.~\ref{fig:CPW_plot}, we show the input excitation (top), and compare the signal evolution at the second location as a function of time as obtained from the four cases.
In each case, the fundamental mode is clearly established in the resonator line, similar to that previously illustrated for the $E_x$ field in Fig.~\ref{fig:CPW_ab}(b).
However, the amplitude of the signal decays rapidly for Case 1, where the superconducting film is approximated as a regular conductor with $\sigma=6\times 10^7$ S/m.
We also observe that the amplitude variation for cases 2, 3, 4, are similar. 
To quantify the difference, we compute the $Q$-factor and resonant frequency obtained from the simulation using the measured signal shown in Fig.~\ref{fig:CPW_plot} \cite{esprit1,esprit2}.
$Q$-factor, also known as quality factor, quantifies how underdamped or performant a resonator is, and is therefore widely used to quantify the efficiency of a resonator.
We compute $Q$-factor beginning the measurement at $t=8\times 10^{-11}$~s, which is after the input pulse has died out, resonance has formed, and at least five periods of resonance are recorded.
We use the signal processing code ESPRIT  \cite{esprit1,esprit2} to extract the attenuation constant and phase constant, which are used to compute the $Q$-factor. 
In Table \ref{tab:qandf0} we report the $Q$-factor and computed resonance frequency, $f_{\rm comp}$, for each of the four cases. 
We see that the standard conductor (Case 1) has the lowest $Q$-factor, and the superconductor modeled with no conductivity (Case 4) has the largest $Q$-factor, i.e., the most performant. 
Cases 2 and 3 show a $Q$-factor that is in-between the purely conductive and the purely superconducting cases.
This indicates that the amount of standard conductivity included in a superconducting model (in physical terms, the temperature of the system) can have a significant impact on performance, and the assumption of quasi-infinite conductivity may not accurately describe performance.
Each simulation is able to compute a resonance frequency that is close to the predicted frequency, consistent with the observation from the time-domain plot show in Fig.~\ref{fig:CPW_plot}. \textcolor{black}{All inputs files, data sets, and scripts used for analysis of the simulations presented in Secs.~\ref{sec:validation} and \ref{sec:coplanar} are provided online \footnote{\url{https://doi.org/10.5281/zenodo.7943012}}}.

\begin{table}[tb!]
\centering
\begin{tabular}{p{2.5cm} p{2.5cm} p{2.5cm}}
\hline
& $Q$ & $f_{\rm comp}$ [GHz] \\
\hline
Case 1 & 24 & 96.9 \\
Case 2 & 491 & 98.2 \\
Case 3 & 316 & 97.1 \\
Case 4 & 973 & 97.1 \\
\hline
\end{tabular}
\caption{Computed $Q$-factors and resonance frequencies for (Case 1) a regular conductor, (Case 2) an artificially-high conductive material, (Case 3) a superconductor with conductivity, and (Case 4) a superconductor with zero conductivity.}
\label{tab:qandf0}
\end{table}

\section{Summary and Future Work}\label{sec:conclusion}
We have implemented a two-fluid model for superconductivity within the open-source ARTEMIS framework and performed numerical studies to validate the model.
We have demonstrated that our algorithm is second-order accurate in space and time within superconducting materials and first-order in space in the presence of superconducting material interfaces.
The reflection coefficient and skin depth obtained from our implementation agree with theoretical predictions for a wide range of material properties and frequencies. 
We have applied our algorithm to model resonant behavior in a superconducting coplanar waveguide, demonstrating that the superconducting physics performs on-par, or even better than the assumption of quasi-infinite conductivity.

There are several avenues to explore for future work in relation to improving our model further and broadening our applications.
To improve the model, we would like to develop an effective absorbing boundary condition along the lines of PML for superconducting interfaces at domain boundaries.
We would like to explore and develop higher-order accurate discretization in space to improve the accuracy of the method at superconducting material interfaces.  
Also, methods that are not subject to the Courant condition such as implicit \cite{yao2018,chen2001} or spectral methods could be explored in order to significantly increase the timestep, which limits the frequency that can be used in 
coplanar waveguide simulations even when using GPUs.
While in this work, we use a constant conductivity and London penetration depth throughout the simulation, new modifications to the model could be implemented to account for the temperature dependence of these quantities using alternate approaches suggested by Hirsch \cite{hirsch2004}. \textcolor{black}{It would also be of interest to explore using a more general Landau-Ginzberg model or an electrodynamic vector potential to compute the superconducting current density\cite{koizumi2020theory} and compare with the two-fluid model implemented in this work.}
For the case of complex geometrical features, e.g., resonator readout circuitry with non-grid aligned transmission lines \textcolor{black}{or spherical/curved geometries where this current work only supports a staircase approximation}, we would like to explore embedded boundary discretizations, which have been developed for Maxwell's equations \cite{benkler2006new,cai2003upwinding,xiao2005enlarged}. Finally, we would like to expand the implementation to applications in larger circuits where we may model multiple superconducting sub-components and develop new methods to quantify crosstalk interactions between them.

\section*{Acknowledgments}
This work was supported by Laboratory Directed Research and Development (LDRD) funding from Berkeley Lab, provided by the Director, Office of Science, of the U.S. Department of Energy under Contract No. DE-AC02-05CH11231.
This work was supported in part by the U.S. Department of Energy, Office of Science, Office of Workforce Development for Teachers and Scientists (WDTS) under the Visiting Faculty Program (VFP).
This research used resources of the National Energy Research Scientific Computing Center (NERSC), a U.S. Department of Energy Office of Science User Facility operated under Contract No. DE-AC02-05CH11231.


\bibliographystyle{elsarticle-num-names}
\bibliography{london.bbl}

\begin{thebibliography}{46}
\expandafter\ifx\csname natexlab\endcsname\relax\def\natexlab#1{#1}\fi
\providecommand{\url}[1]{\texttt{#1}}
\providecommand{\href}[2]{#2}
\providecommand{\path}[1]{#1}
\providecommand{\DOIprefix}{doi:}
\providecommand{\ArXivprefix}{arXiv:}
\providecommand{\URLprefix}{URL: }
\providecommand{\Pubmedprefix}{pmid:}
\providecommand{\doi}[1]{\href{http://dx.doi.org/#1}{\path{#1}}}
\providecommand{\Pubmed}[1]{\href{pmid:#1}{\path{#1}}}
\providecommand{\bibinfo}[2]{#2}
\ifx\xfnm\relax \def\xfnm[#1]{\unskip,\space#1}\fi
\bibitem[{Meisnner and Ochsenfeld(1933)}]{meissner1933}
\bibinfo{author}{W.~Meisnner}, \bibinfo{author}{R.~Ochsenfeld},
\newblock \bibinfo{title}{Ein neuer effekt bei eintritt der
  supraleitfahigkeit},
\newblock \bibinfo{journal}{Naturwissenchaften} \bibinfo{volume}{21}
  (\bibinfo{year}{1933}) \bibinfo{pages}{787}.
\bibitem[{Tinkham(1974)}]{tinkham1974electromagnetic}
\bibinfo{author}{M.~Tinkham},
\newblock \bibinfo{title}{The electromagnetic properties of superconductors},
\newblock \bibinfo{journal}{Reviews of Modern Physics} \bibinfo{volume}{46}
  (\bibinfo{year}{1974}) \bibinfo{pages}{587}.
\bibitem[{Kirtley(2010)}]{kirtley2010fundamental}
\bibinfo{author}{J.~Kirtley},
\newblock \bibinfo{title}{Fundamental studies of superconductors using scanning
  magnetic imaging},
\newblock \bibinfo{journal}{Reports on Progress in Physics}
  \bibinfo{volume}{73} (\bibinfo{year}{2010}) \bibinfo{pages}{126501}.
\bibitem[{Gryaznevich et~al.(2013)Gryaznevich, Svoboda, Stockel, Sykes, Sykes,
  Kingham, Hammond, Apte, Todd, Ball et~al.}]{gryaznevich2013progress}
\bibinfo{author}{M.~Gryaznevich}, \bibinfo{author}{V.~Svoboda},
  \bibinfo{author}{J.~Stockel}, \bibinfo{author}{A.~Sykes},
  \bibinfo{author}{N.~Sykes}, \bibinfo{author}{D.~Kingham},
  \bibinfo{author}{G.~Hammond}, \bibinfo{author}{P.~Apte},
  \bibinfo{author}{T.~Todd}, \bibinfo{author}{S.~Ball}, et~al.,
\newblock \bibinfo{title}{Progress in application of high temperature
  superconductor in tokamak magnets},
\newblock \bibinfo{journal}{Fusion Engineering and Design} \bibinfo{volume}{88}
  (\bibinfo{year}{2013}) \bibinfo{pages}{1593--1596}.
\bibitem[{Rossi and Bottura(2012)}]{rossi2012superconducting}
\bibinfo{author}{L.~Rossi}, \bibinfo{author}{L.~Bottura},
\newblock \bibinfo{title}{Superconducting magnets for particle accelerators},
\newblock \bibinfo{journal}{Reviews of accelerator science and technology}
  \bibinfo{volume}{5} (\bibinfo{year}{2012}) \bibinfo{pages}{51--89}.
\bibitem[{Padamsee(2001)}]{padamsee2001science}
\bibinfo{author}{H.~Padamsee},
\newblock \bibinfo{title}{The science and technology of superconducting
  cavities for accelerators},
\newblock \bibinfo{journal}{Superconductor science and technology}
  \bibinfo{volume}{14} (\bibinfo{year}{2001}) \bibinfo{pages}{R28}.
\bibitem[{Moon(2008)}]{moon2008superconducting}
\bibinfo{author}{F.~C. Moon}, \bibinfo{title}{Superconducting levitation:
  Applications to bearings and magnetic transportation},
  \bibinfo{publisher}{John Wiley \& Sons}, \bibinfo{year}{2008}.
\bibitem[{Ma et~al.(2003)Ma, Postrekhin, and Chu}]{ma2003superconductor}
\bibinfo{author}{K.~Ma}, \bibinfo{author}{Y.~Postrekhin},
  \bibinfo{author}{W.~Chu},
\newblock \bibinfo{title}{Superconductor and magnet levitation devices},
\newblock \bibinfo{journal}{Review of Scientific Instruments}
  \bibinfo{volume}{74} (\bibinfo{year}{2003}) \bibinfo{pages}{4989--5017}.
\bibitem[{Burkard et~al.(2020)Burkard, Gullans, Mi, and
  Petta}]{burkard2020superconductor}
\bibinfo{author}{G.~Burkard}, \bibinfo{author}{M.~J. Gullans},
  \bibinfo{author}{X.~Mi}, \bibinfo{author}{J.~R. Petta},
\newblock \bibinfo{title}{Superconductor--semiconductor hybrid-circuit quantum
  electrodynamics},
\newblock \bibinfo{journal}{Nature Reviews Physics} \bibinfo{volume}{2}
  (\bibinfo{year}{2020}) \bibinfo{pages}{129--140}.
\bibitem[{Wallraff et~al.(2004)Wallraff, Schuster, Blais, Frunzio, Huang,
  Majer, Kumar, Girvin, and Schoelkopf}]{wallraff2004strong}
\bibinfo{author}{A.~Wallraff}, \bibinfo{author}{D.~I. Schuster},
  \bibinfo{author}{A.~Blais}, \bibinfo{author}{L.~Frunzio},
  \bibinfo{author}{R.-S. Huang}, \bibinfo{author}{J.~Majer},
  \bibinfo{author}{S.~Kumar}, \bibinfo{author}{S.~M. Girvin},
  \bibinfo{author}{R.~J. Schoelkopf},
\newblock \bibinfo{title}{Strong coupling of a single photon to a
  superconducting qubit using circuit quantum electrodynamics},
\newblock \bibinfo{journal}{Nature} \bibinfo{volume}{431}
  (\bibinfo{year}{2004}) \bibinfo{pages}{162--167}.
\bibitem[{Caputo et~al.(2021)Caputo, Danaila, and Tain}]{caputo2021abelian}
\bibinfo{author}{J.-G. Caputo}, \bibinfo{author}{I.~Danaila},
  \bibinfo{author}{C.~Tain},
\newblock \bibinfo{title}{An abelian higgs model of pulsed field magnetization
  in superconductors},
\newblock in: \bibinfo{booktitle}{Journal of Physics: Conference Series},
  volume \bibinfo{volume}{2043}, \bibinfo{organization}{IOP Publishing},
  \bibinfo{year}{2021}, p. \bibinfo{pages}{012006}.
\bibitem[{Sage et~al.(2011)Sage, Bolkhovsky, Oliver, Turek, and
  Welander}]{sage2011study}
\bibinfo{author}{J.~M. Sage}, \bibinfo{author}{V.~Bolkhovsky},
  \bibinfo{author}{W.~D. Oliver}, \bibinfo{author}{B.~Turek},
  \bibinfo{author}{P.~B. Welander},
\newblock \bibinfo{title}{Study of loss in superconducting coplanar waveguide
  resonators},
\newblock \bibinfo{journal}{Journal of Applied Physics} \bibinfo{volume}{109}
  (\bibinfo{year}{2011}) \bibinfo{pages}{063915}.
\bibitem[{G{\"o}ppl et~al.(2008)G{\"o}ppl, Fragner, Baur, Bianchetti, Filipp,
  Fink, Leek, Puebla, Steffen, and Wallraff}]{goppl2008coplanar}
\bibinfo{author}{M.~G{\"o}ppl}, \bibinfo{author}{A.~Fragner},
  \bibinfo{author}{M.~Baur}, \bibinfo{author}{R.~Bianchetti},
  \bibinfo{author}{S.~Filipp}, \bibinfo{author}{J.~M. Fink},
  \bibinfo{author}{P.~J. Leek}, \bibinfo{author}{G.~Puebla},
  \bibinfo{author}{L.~Steffen}, \bibinfo{author}{A.~Wallraff},
\newblock \bibinfo{title}{Coplanar waveguide resonators for circuit quantum
  electrodynamics},
\newblock \bibinfo{journal}{Journal of Applied Physics} \bibinfo{volume}{104}
  (\bibinfo{year}{2008}) \bibinfo{pages}{113904}.
\bibitem[{London and London(1935)}]{london1935electromagnetic}
\bibinfo{author}{F.~London}, \bibinfo{author}{H.~London},
\newblock \bibinfo{title}{The electromagnetic equations of the supraconductor},
\newblock \bibinfo{journal}{Proceedings of the Royal Society of London. Series
  A-Mathematical and Physical Sciences} \bibinfo{volume}{149}
  (\bibinfo{year}{1935}) \bibinfo{pages}{71--88}.
\bibitem[{Kunz and Luebbers(1993)}]{kunz1993finite}
\bibinfo{author}{K.~S. Kunz}, \bibinfo{author}{R.~J. Luebbers},
  \bibinfo{title}{The finite difference time domain method for
  electromagnetics}, \bibinfo{publisher}{CRC press}, \bibinfo{year}{1993}.
\bibitem[{Rittweger and Wolff(1992)}]{rittweger1992finite}
\bibinfo{author}{M.~Rittweger}, \bibinfo{author}{I.~Wolff},
\newblock \bibinfo{title}{Finite difference time-domain formulation for
  transient propagation in superconductors},
\newblock in: \bibinfo{booktitle}{IEEE Antennas and Propagation Society
  International Symposium 1992 Digest}, \bibinfo{organization}{IEEE},
  \bibinfo{year}{1992}, pp. \bibinfo{pages}{1960--1963}.
\bibitem[{Xiao and Vahldieck(1994)}]{xiao1994extended}
\bibinfo{author}{S.~Xiao}, \bibinfo{author}{R.~Vahldieck},
\newblock \bibinfo{title}{An extended 2d fdtd method for hybrid mode analysis
  of lossy and superconducting structures},
\newblock in: \bibinfo{booktitle}{Proceedings of IEEE Antennas and Propagation
  Society International Symposium and URSI National Radio Science Meeting},
  volume~\bibinfo{volume}{3}, \bibinfo{organization}{IEEE},
  \bibinfo{year}{1994}, pp. \bibinfo{pages}{1774--1777}.
\bibitem[{Xiao and Vahldieck(1995)}]{xiao19953d}
\bibinfo{author}{S.~Xiao}, \bibinfo{author}{R.~Vahldieck},
\newblock \bibinfo{title}{3d fdtd simulation of superconductor coplanar
  waveguides},
\newblock in: \bibinfo{booktitle}{Proceedings of 1995 IEEE MTT-S International
  Microwave Symposium}, \bibinfo{organization}{IEEE}, \bibinfo{year}{1995}, pp.
  \bibinfo{pages}{349--352}.
\bibitem[{Hofschen and Wolff(1996)}]{hofschen1996improvements}
\bibinfo{author}{S.~Hofschen}, \bibinfo{author}{I.~Wolff},
\newblock \bibinfo{title}{Improvements of the two-dimensional fdtd method for
  the simulation of normal-and superconducting planar waveguides using time
  series analysis},
\newblock \bibinfo{journal}{IEEE transactions on microwave theory and
  techniques} \bibinfo{volume}{44} (\bibinfo{year}{1996})
  \bibinfo{pages}{1487--1490}.
\bibitem[{Gorter and Casimir(1934)}]{Gorter34}
\bibinfo{author}{C.~J. Gorter}, \bibinfo{author}{H.~Casimir},
\newblock \bibinfo{title}{On supraconductivity i},
\newblock \bibinfo{journal}{Physica} \bibinfo{volume}{1} (\bibinfo{year}{1934})
  \bibinfo{pages}{306--320}.
\bibitem[{Van~Duzer and Turner(1981)}]{van1981principles}
\bibinfo{author}{T.~Van~Duzer}, \bibinfo{author}{C.~W. Turner},
  \bibinfo{title}{Principles of superconductive devices and circuits},
  \bibinfo{publisher}{Prentice Hall}, \bibinfo{year}{1981}.
\bibitem[{Megahed and El-Ghazaly(1995)}]{megahed1995nonlinear}
\bibinfo{author}{M.~A. Megahed}, \bibinfo{author}{S.~M. El-Ghazaly},
\newblock \bibinfo{title}{Nonlinear analysis of microwave superconductor
  devices using full-wave electromagnetic model},
\newblock \bibinfo{journal}{IEEE Transactions on Microwave Theory and
  Techniques} \bibinfo{volume}{43} (\bibinfo{year}{1995})
  \bibinfo{pages}{2590--2599}.
\bibitem[{Okazaki et~al.(1999)Okazaki, Suzuki, and
  Enomoto}]{okazaki1999superconducting}
\bibinfo{author}{Y.~Okazaki}, \bibinfo{author}{K.~Suzuki},
  \bibinfo{author}{Y.~Enomoto},
\newblock \bibinfo{title}{Superconducting microstrip resonator investigated by
  fdtd electromagnetic field simulator},
\newblock \bibinfo{journal}{IEEE Transactions on applied superconductivity}
  \bibinfo{volume}{9} (\bibinfo{year}{1999}) \bibinfo{pages}{3034--3037}.
\bibitem[{Mollai et~al.(2013)Mollai, Javadzadeh, Shishegar, Banai, Farzaneh,
  and Fardmanesh}]{mollai2013analysis}
\bibinfo{author}{S.~Mollai}, \bibinfo{author}{S.~M.~H. Javadzadeh},
  \bibinfo{author}{A.~A. Shishegar}, \bibinfo{author}{A.~Banai},
  \bibinfo{author}{F.~Farzaneh}, \bibinfo{author}{M.~Fardmanesh},
\newblock \bibinfo{title}{Analysis of nonlinearities in superconducting
  microstrip straight bends; fdtd method in comparison with nonlinear circuit
  modeling},
\newblock \bibinfo{journal}{Journal of superconductivity and novel magnetism}
  \bibinfo{volume}{26} (\bibinfo{year}{2013}) \bibinfo{pages}{1827--1830}.
\bibitem[{Li et~al.(2021)Li, Li, Yang, and Mao}]{li2021unconditionally}
\bibinfo{author}{Y.~Li}, \bibinfo{author}{X.-C. Li}, \bibinfo{author}{Y.~Yang},
  \bibinfo{author}{J.-F. Mao},
\newblock \bibinfo{title}{An unconditionally stable 2-d stochastic wlp-fdtd
  method for geometric uncertainty in superconducting transmission lines},
\newblock \bibinfo{journal}{IEEE Transactions on Applied Superconductivity}
  \bibinfo{volume}{32} (\bibinfo{year}{2021}) \bibinfo{pages}{1--9}.
\bibitem[{Li et~al.(2022)Li, Li, and Mao}]{li2022extended}
\bibinfo{author}{Y.~Li}, \bibinfo{author}{X.-C. Li}, \bibinfo{author}{J.-F.
  Mao},
\newblock \bibinfo{title}{Extended fdtd method for co-analysis of
  superconducting passive transmission lines and josephson junction drivers and
  receivers},
\newblock \bibinfo{journal}{IEEE Transactions on Applied Superconductivity}
  (\bibinfo{year}{2022}).
\bibitem[{Shen-Yun et~al.(2010)Shen-Yun, Shao-Bin, and Li}]{Shen-Yun10}
\bibinfo{author}{W.~Shen-Yun}, \bibinfo{author}{L.~Shao-Bin},
  \bibinfo{author}{L.-W.~J. Li},
\newblock \bibinfo{title}{Finite-difference time-domain studies of
  low-frequency stop band in superconductor-dielectric superlattice},
\newblock \bibinfo{journal}{Chinese Physics B} \bibinfo{volume}{19}
  (\bibinfo{year}{2010}) \bibinfo{pages}{084101}.
\bibitem[{Yao et~al.(2022)Yao, Jambunathan, Zeng, and
  Nonaka}]{yao2022massively}
\bibinfo{author}{Z.~Yao}, \bibinfo{author}{R.~Jambunathan},
  \bibinfo{author}{Y.~Zeng}, \bibinfo{author}{A.~Nonaka},
\newblock \bibinfo{title}{A massively parallel time-domain coupled
  electrodynamics--micromagnetics solver},
\newblock \bibinfo{journal}{The International Journal of High Performance
  Computing Applications} \bibinfo{volume}{36} (\bibinfo{year}{2022})
  \bibinfo{pages}{167--181}.
\bibitem[{Bardeen et~al.(1957)Bardeen, Cooper, and Schrieffer}]{Bardeen57}
\bibinfo{author}{J.~Bardeen}, \bibinfo{author}{L.~N. Cooper},
  \bibinfo{author}{J.~R. Schrieffer},
\newblock \bibinfo{title}{Theory of superconductivity},
\newblock \bibinfo{journal}{Physical Review} \bibinfo{volume}{108}
  (\bibinfo{year}{1957}) \bibinfo{pages}{1175--1204}.
\bibitem[{Berenger(1996)}]{berenger1996three}
\bibinfo{author}{J.-P. Berenger},
\newblock \bibinfo{title}{Three-dimensional perfectly matched layer for the
  absorption of electromagnetic waves},
\newblock \bibinfo{journal}{Journal of Computational Physics}
  \bibinfo{volume}{127} (\bibinfo{year}{1996}) \bibinfo{pages}{363--379}.
\bibitem[{Shapoval et~al.(2019)Shapoval, Vay, and Vincenti}]{shapoval2019two}
\bibinfo{author}{O.~Shapoval}, \bibinfo{author}{J.-L. Vay},
  \bibinfo{author}{H.~Vincenti},
\newblock \bibinfo{title}{Two-step perfectly matched layer for arbitrary-order
  pseudo-spectral analytical time-domain methods},
\newblock \bibinfo{journal}{Computer Physics Communications}
  \bibinfo{volume}{235} (\bibinfo{year}{2019}) \bibinfo{pages}{102--110}.
\bibitem[{Zhang et~al.(2021)Zhang, Myers, Gott, Almgren, and
  Bell}]{zhang2021amrex}
\bibinfo{author}{W.~Zhang}, \bibinfo{author}{A.~Myers},
  \bibinfo{author}{K.~Gott}, \bibinfo{author}{A.~Almgren},
  \bibinfo{author}{J.~Bell},
\newblock \bibinfo{title}{{AMReX}: Block-structured adaptive mesh refinement
  for multiphysics applications},
\newblock \bibinfo{journal}{The International Journal of High Performance
  Computing Applications} \bibinfo{volume}{35} (\bibinfo{year}{2021})
  \bibinfo{pages}{508--526}.
\bibitem[{Vay et~al.(2018)Vay, Almgren, Bell, Ge, Grote, Hogan, Kononenko,
  Lehe, Myers, Ng et~al.}]{vay2018warp}
\bibinfo{author}{J.-L. Vay}, \bibinfo{author}{A.~Almgren},
  \bibinfo{author}{J.~Bell}, \bibinfo{author}{L.~Ge},
  \bibinfo{author}{D.~Grote}, \bibinfo{author}{M.~Hogan},
  \bibinfo{author}{O.~Kononenko}, \bibinfo{author}{R.~Lehe},
  \bibinfo{author}{A.~Myers}, \bibinfo{author}{C.~Ng}, et~al.,
\newblock \bibinfo{title}{{Warp-X: A new exascale computing platform for
  beam--plasma simulations}},
\newblock \bibinfo{journal}{Nuclear Instruments and Methods in Physics Research
  Section A: Accelerators, Spectrometers, Detectors and Associated Equipment}
  \bibinfo{volume}{909} (\bibinfo{year}{2018}) \bibinfo{pages}{476--479}.
\bibitem[{Sawant et~al.(2022)Sawant, Yao, Jambunathan, and Nonaka}]{Sawant2022}
\bibinfo{author}{S.~S. Sawant}, \bibinfo{author}{Z.~Yao},
  \bibinfo{author}{R.~Jambunathan}, \bibinfo{author}{A.~Nonaka},
\newblock \bibinfo{title}{Characterization of transmission lines in
  microelectronic circuits using the artemis solver},
\newblock \bibinfo{journal}{IEEE Journal on Multiscale and Multiphysics
  Computational Techniques} \bibinfo{volume}{8} (\bibinfo{year}{2022})
  \bibinfo{pages}{31--39}.
\bibitem[{Griffiths(1981)}]{griffiths2005introduction}
\bibinfo{author}{D.~J. Griffiths}, \bibinfo{title}{Introduction to
  electrodynamics}, \bibinfo{publisher}{Prencie-Hall}, \bibinfo{year}{1981}.
\bibitem[{Chen and Chou(1997)}]{chen1997characteristics}
\bibinfo{author}{E.~Chen}, \bibinfo{author}{S.~Y. Chou},
\newblock \bibinfo{title}{Characteristics of coplanar transmission lines on
  multilayer substrates: Modeling and experiments},
\newblock \bibinfo{journal}{IEEE transactions on microwave theory and
  techniques} \bibinfo{volume}{45} (\bibinfo{year}{1997})
  \bibinfo{pages}{939--945}.
\bibitem[{Berenger et~al.(1994)}]{berenger1994perfectly}
\bibinfo{author}{J.-P. Berenger}, et~al.,
\newblock \bibinfo{title}{A perfectly matched layer for the absorption of
  electromagnetic waves},
\newblock \bibinfo{journal}{Journal of Computational Physics}
  \bibinfo{volume}{114} (\bibinfo{year}{1994}) \bibinfo{pages}{185--200}.
\bibitem[{{Roy} and {Kailath}(1989)}]{esprit1}
\bibinfo{author}{R.~{Roy}}, \bibinfo{author}{T.~{Kailath}},
\newblock \bibinfo{title}{Esprit-estimation of signal parameters via rotational
  invariance techniques},
\newblock \bibinfo{journal}{IEEE Transactions on Acoustics, Speech, and Signal
  Processing} \bibinfo{volume}{37} (\bibinfo{year}{1989})
  \bibinfo{pages}{984--995}. \DOIprefix\doi{10.1109/29.32276}.
\bibitem[{{Wang} and {Ling}(1998)}]{esprit2}
\bibinfo{author}{Y.~{Wang}}, \bibinfo{author}{H.~{Ling}},
\newblock \bibinfo{title}{Multimode parameter extraction for multiconductor
  transmission lines via single-pass fdtd and signal-processing techniques},
\newblock \bibinfo{journal}{IEEE Transactions on Microwave Theory and
  Techniques} \bibinfo{volume}{46} (\bibinfo{year}{1998})
  \bibinfo{pages}{89--96}. \DOIprefix\doi{10.1109/22.654927}.
\bibitem[{Yao et~al.(2018)Yao, Tok, Itoh, and Wang}]{yao2018}
\bibinfo{author}{Z.~Yao}, \bibinfo{author}{R.~U. Tok},
  \bibinfo{author}{T.~Itoh}, \bibinfo{author}{Y.~E. Wang},
\newblock \bibinfo{title}{A multiscale unconditionally stable time-domain
  (must) solver unifying electrodynamics and micromagnetics},
\newblock \bibinfo{journal}{IEEE Transactions on Microwave Theory and
  Techniques} \bibinfo{volume}{66} (\bibinfo{year}{2018})
  \bibinfo{pages}{2683--2696}. \DOIprefix\doi{10.1109/TMTT.2018.2825373}.
\bibitem[{Zheng and Chen(2001)}]{chen2001}
\bibinfo{author}{F.~Zheng}, \bibinfo{author}{Z.~Chen},
\newblock \bibinfo{title}{Numerical dispersion analysis of the unconditionally
  stable 3-d adi-fdtd method},
\newblock \bibinfo{journal}{IEEE Transactions on Microwave Theory and
  Techniques} \bibinfo{volume}{49} (\bibinfo{year}{2001})
  \bibinfo{pages}{1006--1009}. \DOIprefix\doi{10.1109/22.920165}.
\bibitem[{Hirsch(2004)}]{hirsch2004}
\bibinfo{author}{J.~E. Hirsch},
\newblock \bibinfo{title}{Electrodynamics of superconductors},
\newblock \bibinfo{journal}{Physical Review B} \bibinfo{volume}{69}
  (\bibinfo{year}{2004}) \bibinfo{pages}{214515}.
\bibitem[{Koizumi and Ishikawa(2020)}]{koizumi2020theory}
\bibinfo{author}{H.~Koizumi}, \bibinfo{author}{A.~Ishikawa},
\newblock \bibinfo{title}{Theory of supercurrent in superconductors},
\newblock \bibinfo{journal}{International Journal of Modern Physics B}
  \bibinfo{volume}{34} (\bibinfo{year}{2020}) \bibinfo{pages}{2030001}.
\bibitem[{Benkler et~al.(2006)Benkler, Chavannes, and Kuster}]{benkler2006new}
\bibinfo{author}{S.~Benkler}, \bibinfo{author}{N.~Chavannes},
  \bibinfo{author}{N.~Kuster},
\newblock \bibinfo{title}{A new 3-d conformal pec fdtd scheme with user-defined
  geometric precision and derived stability criterion},
\newblock \bibinfo{journal}{IEEE Transactions on Antennas and Propagation}
  \bibinfo{volume}{54} (\bibinfo{year}{2006}) \bibinfo{pages}{1843--1849}.
\bibitem[{Cai and Deng(2003)}]{cai2003upwinding}
\bibinfo{author}{W.~Cai}, \bibinfo{author}{S.~Deng},
\newblock \bibinfo{title}{An upwinding embedded boundary method for maxwell’s
  equations in media with material interfaces: 2d case},
\newblock \bibinfo{journal}{Journal of Computational Physics}
  \bibinfo{volume}{190} (\bibinfo{year}{2003}) \bibinfo{pages}{159--183}.
\bibitem[{Xiao and Liu(2005)}]{xiao2005enlarged}
\bibinfo{author}{T.~Xiao}, \bibinfo{author}{Q.~Liu},
\newblock \bibinfo{title}{An enlarged cell technique for the conformal fdtd
  method to model perfectly conducting objects},
\newblock in: \bibinfo{booktitle}{2005 IEEE Antennas and Propagation Society
  International Symposium}, volume~\bibinfo{volume}{1},
  \bibinfo{organization}{IEEE}, \bibinfo{year}{2005}, pp.
  \bibinfo{pages}{122--125}.

\end{thebibliography}

\end{document}